\documentclass[11pt]{article}
\pdfoutput=1
\usepackage{latexsym}
\usepackage{amssymb}
\usepackage{epsf}
\usepackage{epsfig}
\usepackage{amsmath}
\usepackage[bookmarks=true]{hyperref}
\usepackage{graphicx}
\usepackage{amscd}
\usepackage{comment}
\usepackage[usenames]{color}

\newcommand{\tr}{{\rm Tr}}

\begin{document}
\pagestyle{empty}

\begin{flushright}
DESY 11-053 \\
TU-881 \\
\today 
\end{flushright}

\vspace{3cm}

\begin{center}

{\bf\LARGE On Supersymmetric Effective Theories of Axion}
\\

\vspace*{1.5cm}
{\large 
Tetsutaro Higaki$^1$ and Ryuichiro Kitano$^2$
} \\
\vspace*{0.5cm}

$^1${\it DESY Theory Group, Notkestrasse 85, D-22607 Hamburg, Germany}\\
\vspace*{0.5cm}

$^2${\it Department of Physics, Tohoku University, Sendai 980-8578, Japan}\\
\vspace*{0.5cm}

\end{center}

\vspace*{1.0cm}

\begin{abstract}
{\normalsize
We study effective theories of an axion in spontaneously broken
supersymmetric theories. We consider a system where the axion
supermultiplet is directly coupled to a supersymmetry breaking sector
whereas the standard model sector is communicated with those sectors
through loops of messenger fields. The gaugino masses and the
axion-gluon coupling necessary for solving the strong CP problem are
both obtained by the same effective interaction. We discuss cosmological
constraints on this framework.}
%
%
\end{abstract} 

\newpage
\baselineskip=18pt
\setcounter{page}{2}
\pagestyle{plain}

\setcounter{footnote}{0}

\section{Introduction}

The hierarchy problem~\cite{Weinberg:1979bn} and the strong CP
problem~\cite{Callan:1976je} are two big misteries unexplained in the
Standard Model of particle physics.
Among various candidate solutions proposed so far, supersymmetry (SUSY)
and the Peccei-Quinn (PQ)~\cite{Peccei:1977hh} mechanism are promissing
ones. Both hypotheses rely on broken symmetries. The construction of the
whole picture requires model building efforts of the symmetry breaking
sectors and the mediation mechanisms to the Standard Model sector.

The consistency needs to be checked when we combine SUSY and the PQ
mechanism.
If we are meant to solve the strong CP problem, it is not a good idea to
assume a mechanism which causes another CP problem such as the SUSY CP
problem. This gives a constraint on the SUSY breaking/mediation
sector. Also, by SUSY the axion field is necessarily extended to an
axion supermultiplet, which includes a CP-even axion field (saxion). The
field value of the saxion (corresponding to the axion decay constant)
needs to be stabilized, that requries a coupling to the SUSY breaking
sector. The decay constant $f_a$ is constrained by cosmology and
astrophysics. The viable region, $10^9~{\rm GeV} \lesssim f_a \lesssim
10^{12}$ GeV, is called the axion window \cite{Janka:1995ir}.

There are numbers of approaches for a supersymmetric axion model in field
theories and string theories~\cite{Kim:1983dt}-\cite{Lee:2011dy}.
%
%
The consistent model building tends to be rather complicated because one
needs to carefully glue two symmetry breaking sectors. Especially, field
theoretic approaches typically find new particles in both the PQ
breaking sector and the SUSY breaking sector. The discussion of the
viability is limited to specific models in such cases.
On the other hand, the string theory approach tends to predict a too
large decay constant \cite{Choi:1985je} for the PQ symmetry breaking, 
such as the Planck scale, since that is the only scale in the theory unless
there is some non-perturbative effects or a large compactification
volume.
The cosmological difficulty of the saxion (or moduli) field has also
been pointed out~\cite{Banks:2002sd, Polonyi:1977pj, Coughlan:1983ci}.
The coherent oscillation of the saxion field and its decay produces
dangerous particles like gravitino. There are severe constraints from
the Big Bang nucleosynthesis (BBN)~\cite{Kawasaki:2004yh,
Kawasaki:2008qe} and also the matter density of the
universe~\cite{Komatsu:2010fb} if the gravitino is the lightest
supersymmetric particle (LSP).


In order to avoid the complexity of the models for the discussion, we
study simplified models of SUSY and PQ breaking in this paper. We
construct effective theories which cover various microscopic models,
although we do make a few assumptions motivated by
phenomenology/cosmology and minimality. First, we assume the gauge
mediation mechanism for the transmission of the SUSY breaking. This is a
good starting point to avoid the SUSY FCNC and CP problem. We assume a
direct coupling between the SUSY and the PQ breaking sectors so that the
saxion can obtain a large enough mass to avoid the cosmological
difficulty. Finally, we assume that the same messenger fields (or the
same effective operator) to communicate the SUSY breaking and PQ
breaking to the Standard Model sector.

In the next Section, we will propose a general framework for the
simplified model. In Section~3, we will study properties of three
representative models classfied by ways to break an $R$-symmetry. The
cosmological constraints are discussed in Section~4.
%
In Section~5, we will give brief comments on the string theory approach.
Section~6 will be devoted to the conclusion.  In the appendices, we
discuss supergravity corrections to the mass spectrum and list relevant
computations of the decay rates of heavy fields.


\section{Low energy effective Lagrangian}

We construct a model where the PQ symmetry is non-linearly realized
whereas SUSY is linearly realized. We introduce an axion chiral
superfield $A$ and a SUSY breaking chiral superfield $X$ whose
$F$-component obtains a non-vanising vacuum expectation value (vev).
%
%
In general, $X$ can carry a PQ charge $q_X$. The superfields $X$ and $A$
transform under the global $U(1)_{\rm PQ}$ symmetry:
\begin{eqnarray}
X \to e^{i q_X \theta_{\rm PQ}}X, \ \ \ 
A \to A + i\theta_{\rm PQ} .
\end{eqnarray}
Here $\theta_{\rm PQ}$ is a transformation parameter.  Without a loss of
generality, we take $q_X \geq 0$.
%

The effective theory for the PQ breaking sector is given by
\begin{eqnarray}
{ K_{\rm axion}} =  
f_0^2 \left[
{1 \over 2} (A + A^\dagger)^2 + \frac{\hat{C}_3}{3!} (A+A^{\dag})^3 +\cdots 
\right].
\end{eqnarray}
No superpotential cannot be written only with $A$. The parameter $f_0$ is
the decay constant of the PQ symmetry breaking. We will take the coefficient $\hat{C}_3$ as
a dimensionless parameter of $O(1)$.
The SUSY breaking sector
has the K{\" a}hler potential:
\begin{eqnarray}
 K_{\rm SB} = \Lambda^2_0 \left[
X^\dagger X - {a (X^\dagger X)^2 \over 4 } - \frac{(X^\dagger X)^3}{18
\hat{x}_0^2} + \cdots 
\right],
\end{eqnarray}
%
where $\Lambda_0$ is the typical mass scale of the SUSY breaking sector
and the parameter $a$ can be chosen to be $a=\pm 1$ by an rescaling of
the field. $\hat x_0^2$ is a prameter of order unity.  The sign of $a$
controls whether $X$ obtains a vev~\cite{Shih:2007av}.
The $R$-symmetry is assumed here so that the SUSY breaking model
described below will be justified.
In the above K{\" a}hler potentials, we took $A$ and $X$ be
dimensionless. Note that their values should be limited to $|X| < 1$ and
$|A| < 1$ for the validity of the effective theory\footnote{In order for
$X$ vev not to dominate the axion decay constant, it is required that
$f_0/\Lambda_0 > \langle X \rangle$ if $q_X \neq 0$.}.

The SUSY breaking is described by the Polonyi
model~\cite{Polonyi:1977pj, Izawa:1996pk, Intriligator:2006dd}. The
superpotential is~\cite{Dudas:2007nz, Heckman:2008es, Heckman:2008qt}:
\begin{eqnarray}
 W_{\rm SB}& = \mu_0^2 ( W_0 +  \Lambda_0 X e^{-q_X A} ).
\label{eq:super}
\end{eqnarray}
Here the $X$ superfield has an $R$-charge 2. The first term is a
constant which is related to a fine-tuning of the cosmological constant
in supergravity, $W_0 \simeq M_{\rm Pl}/\sqrt 3$. The gravitino mass is
$\mu_0^2 W_0 \simeq m_{3/2}M_{\rm Pl}^2$. Here $M_{\rm Pl} = 2.4 \times 10^{18}$ GeV is the reduced Planck mass.
Note that $A \to \infty$ can
be the SUSY vacuum; in that case a minimum with a finite $\langle A
\rangle$ is a metastable SUSY breaking vacuum.


Finally, we write down a coupling between the $X$ and $A$ fields which
is necessary for stabilizaing the saxion field:
\begin{eqnarray}
{K_{AX} } &=& 
\Lambda_{c,0}^2 \Bigg[
\left( c_1 (A+A^\dagger)
+ { c_2 \over 2} (A+A^\dagger)^2 \right) X^\dagger  X 
\\
&& 
\ \ \ 
- d_1 (A+A^{\dag})\frac{a (X^\dag X)^2}{4} -
 {e_1}(A+A^{\dag})\frac{(X^\dag X)^3}{18\hat{x}_0^2}
+\cdots
\Bigg].
\end{eqnarray}
Again, $c_1$, $c_2$, $d_1$ and $e_1$ are parameters of order unity. A
negative value of $c_2$ give a mass term for the saxion. The overall
scale is set by
\begin{eqnarray}
\Lambda_{c,0} \equiv {\rm min}[\Lambda_0 , f_0].
\label{eq:scale}
\end{eqnarray}
The choice of this overall scale is motivated from the discussion below.
%
%
%
%

For $\Lambda_0 \ll f_0$, the above interaction terms are generated
through, for example, a loop of a heavy field in the SUSY breaking
sector whose mass carries the PQ charge\footnote{We can consider a
O'Raifeartaigh model \cite{O'Raifeartaigh:1975pr} in the UV scale such
like $W_{O'} = \mu^2 e^{-q_X A}X + \frac{\Lambda_0}{4\pi} e^{q A} \phi_1
\phi_2 + \tilde\lambda e^{q' A} X \phi_1^2 $.}
\begin{eqnarray}
\Lambda_0^2 X^\dagger X \log | \Lambda_0 e^{q A} |^2.
\end{eqnarray}
This corresponds to above interaction terms with $\Lambda_{c,0} \sim \Lambda_0$.
%

For $\Lambda_0 \gg f_0$, on the other hand, the following terms can be
generated after integrating out heavy modes in the SUSY breaking sector:
\begin{eqnarray}
{S^\dag S}{X^{\dag}X},
\end{eqnarray}
where 
$S = f_0
e^{A}$. This case corresponds to $\Lambda_{c,0} \sim f_0$. 
It is possible that the loops of fields in the PQ breaking sector
generate $1/f_0^2$ suppressed terms which connect $X$ and $A$. In that
case, it is natural to assume that the $|X|^4$ term is also generated
with the same suppression factor. Therefore, $\Lambda_{c,0} \sim \Lambda_0
\sim f_0$.
In any case, one can summarize the overall scale as in
Eq.~(\ref{eq:scale}).


In summary the effective theory we consider is
\begin{eqnarray}
\nonumber
K &=& K_{\rm axion} +K_{\rm SB} +K_{AX} , \\
W &=& W_{\rm SB} .
\label{effL}
\end{eqnarray}
We take all the dimensionless parameters to be of $O(1)$.



\subsection{Unified origins of the axion-gluon coupling and the gaugino mass}

In this subsection, let us consider the origin of an axion coupling to
gluons:
\begin{eqnarray}
\int d^2\theta A {\rm Tr}W^{\alpha}W_{\alpha},
\end{eqnarray}
where $W^\alpha$ is the gluon superfield. This term is necessary to
solve the strong CP problem.
Once $F$-component of $A$ acquires a vev, the same interaction induces
the gluino mass.
Since the term $ X {\rm Tr}W^{\alpha}W_{\alpha}$ is forbidden by
the $R$-symmetry, the above interaction term is going to be the leading
contribution to the gluino mass.
%


This unification of the axion coupling and gaugino masses can often be
seen in string models (see \cite{Conlon:2006tq} for the string theoretic QCD axion with intermediate scale decay constant\footnote{
See also topics
related with LARGE volume scenario
\cite{Balasubramanian:2005zx, Conlon:2007gk, Choi:2010gm}.}).
There, we have a coupling at tree level
\begin{eqnarray}
\frac{1}{2}\int d^2\theta \, \left(\frac{1}{g_h^2}+A \right) \, 
{\tr}W^{\alpha}W_{\alpha} +c.c. 
\end{eqnarray}
Here ${1}/{g_h^2} = {1}/{g_0^2}-i{\vartheta}/{8\pi^2}$.  Now ${\rm
Im}(A)$ can be identified with (a linear combination of) the QCD axion
and $({g_0^2}/{2}) F^A$ is a gaugino mass, which can be comparable to or
less than the gravitino mass.


In gauge mediation models~\cite{Dine:1981za, Dine:1981gu, Dine:1993yw,
Giudice:1998bp}, the above coupling can be obtained after integrating
out messenger fields whose mass carries a PQ charge\footnote{ 
Even though we have an additional $R$-breaking operator
$W = \lambda ' \Lambda X e^{-(q_X + q_{\Psi \bar \Psi})A}\Psi \bar \Psi $,
it is irrelevant to mass spectra, so far as the following conditions are satisfied: 
$M_0 > \lambda' \Lambda Xe^{-q_X A}$ and $M_0 F^A > \lambda' \Lambda F^X
e^{-q_X A}$.
},
\begin{eqnarray}
W_{\rm Mess} = M_0 e^{-q_{\Psi \bar \Psi}A}\Psi \bar \Psi , 
\end{eqnarray}
where $q_{\Psi \bar \Psi} \equiv PQ(\Psi) + PQ(\bar\Psi) $ and $\Psi$
and $\bar \Psi$ are messenger fields.
%
After integrating out the messenger fields, we have the axion coupling
to gauge multiplets,
\begin{eqnarray}
\frac{1}{2}\int d^2\theta \, \left(\frac{1}{g_h^2}
- q_{\Psi \bar \Psi} N_{\Psi \bar \Psi} {A \over 8 \pi^2} \right) \, 
{\tr}W^{\alpha}W_{\alpha} +c.c. 
%
\end{eqnarray}
This is a hadronic axion model~\cite{Kim:1979if}.
Here $N_{\Psi \bar \Psi}$ is the number of messenger
multiplets\footnote{Once we require $q_{\Psi \bar \Psi} N_{\Psi \bar
\Psi} $ should be an integer, a value that $q_{\Psi \bar \Psi} N_{\Psi
\bar \Psi} = 1$ is desirable considering the domain wall
problem \cite{Sikivie:1982qv, Kim:1986ax}}.
%
%
%
${\rm Im}(A)$ is the QCD axion and $q_{\Psi \bar \Psi} N_{\Psi \bar
\Psi} ({\alpha}/{4\pi}) F^A$ is the gaugino mass, which can be larger
than the gravitino mass.

\subsection{Standard Form}

We now define a particular basis to proceed the discussion. First, one
can eliminate $A$ in the superpotential by a redefinition of $X$ by
\begin{eqnarray}
 X \to e^{q_X  A } X.
\end{eqnarray}
%
Next, we can also eliminate the $(A+A^\dagger)X^\dagger X$ term by an
appropriate shift of $A$.
%
The vev $\langle A \rangle$ in this basis is vanishing up to small
$R$-breaking effects we discuss later.
With new definitions of parameters, the superpotential and the K{\"
a}hler potential are given by
\begin{eqnarray}
{W } = \mu^2 ( W_0 + \Lambda X )  ,
\label{standardw}
\end{eqnarray}
\begin{eqnarray}
K &=& K_{\rm axion}  + K_{\rm SB} + K_{AX} ,
\label{standardk}
\end{eqnarray}
where
\begin{eqnarray}
{K_{\rm axion}} = 
f^2 \left[
{1 \over 2} (A + A^\dagger )^2 + { C_3 \over 3!} (A + A^\dagger )^3
+ \cdots
\right],
\end{eqnarray}
\begin{eqnarray}
{K_{\rm SB}} =
\Lambda^2 \left[
   X^\dagger X 
- {a (X^\dagger X)^2 \over 4 } - { (X^\dagger X)^3 \over 18 x_0^2}  
+ \cdots
\right],
\end{eqnarray}
\begin{eqnarray}
{K_{AX} } =
\Lambda_c^2 \left[
 - {\tilde c \over 2}  (A + A^\dagger)^2 X^\dagger X  
- \tilde d (A+A^{\dag}){ a (X^\dagger X)^2 \over 4 }
- \tilde e(A+A^{\dag}){ (X^\dagger X)^3 \over 18 x_0^2}
+ \cdots
\right],
\end{eqnarray}
and
\begin{eqnarray}
 \Lambda_c = {\rm min} [\Lambda, f].
\end{eqnarray}

The gauge kinetic term is 
\begin{eqnarray}
 \int d^2 \theta 
\left( {1 \over g_h^2} + {k \over 8 \pi^2} A \right)
W^\alpha W_\alpha.
\end{eqnarray}
We introduced a parameter $k$ that depends on the messenger mechanism.

One needs to check if the shift we performed is within the allowed range
in the effective theory. In terms of the original parameters, the shift
is given by
\begin{eqnarray}
 \langle A_0 \rangle = - {c_1 \over 2 c_2} 
- \frac{1}{2}\left(
{{1 \over q_X} - \xi}
\right),
\label{vevA}
\end{eqnarray}
where
\begin{eqnarray}
 \xi = \sqrt{{\left(\frac{c_1}{c_2}\right)^2 
+ {1 \over q_X^2} 
-  \frac{2}{c_2} 
\left( {\Lambda_{0} \over \Lambda_{c,0} } \right)^2
}} .
\end{eqnarray}
For $|\langle A_0 \rangle | \lesssim 1$, we need $q_X = 0$ for
$\Lambda \gg f$.

\section{$R$-breaking Models}

The scalar potential can simply be calculated from (\ref{standardw}) and
(\ref{standardk}):
\begin{eqnarray}
V = K^{X^{\dag}X} |\partial_X W|^2.
\end{eqnarray}
Recall that a small $R$-symmetry breaking effect is required to produce
gaugino mass; a vev of $F^A$ is vanishing without $R$-breaking
effects.

We consider three types of $R$-symmetry breaking in the following. The
first one (Model 1) is a model with spontaneous $R$-symmetry
breaking, that can be achieved by taking the $a$ parameter to be $-1$.
The second model (Model 2) is to add a small explicit $R$-breaking
term in the K{\" a}hler potential:
\begin{eqnarray}
 {\Delta K} = \epsilon_K \Lambda^2 (X+X^{\dag})(A+A^{\dag}).
\end{eqnarray}
Finally, the third model (Model 3) is to add a small explicit
$R$-breaking term in the superpotential\footnote{
$\Delta K$ can be, for example, generated from an additional $R$-breaking messenger term
$W= \lambda ' \Lambda X e^{-(q_X + q_{\Psi \bar \Psi})A}\Psi \bar \Psi$ \cite{Cheung:2010jx}.  
On the other hand, $\Delta W$ can originate from a non-renormalizable superpotential $W =
\frac{\Lambda^2}{m} s^2 X^2$, where $s=f e^{-q_X A}$. 
}:
\begin{eqnarray}
 {\Delta W} = \epsilon \mu^2 \Lambda \cdot \frac{X^2}{2} .
\end{eqnarray}
We assume $a=1$ in Model 2 and 3.

We discuss in the following the mass spectrum in each model.

\subsection{Model 0}

Before going to the discussion of the $R$-breaking models listed above,
we comment on the effect of gravity mediation. 
Since the $R$-symmetry is explicitly broken in the supergravity
Lagrangian, $F^A$ is induced and the gauginos obtain masses of
$O(m_{3/2})$.

The axino (the fermionic component of $A$) and the saxion ($A_R$) can
also obtain masses. The axino mass is of $O(m_{3/2})$ and
\begin{eqnarray}
m_{A_R} = {\Lambda_c \over \Lambda} 
\sqrt{6\tilde c}\frac{m_{3/2}M_{\rm Pl}}{f}.
\end{eqnarray}
The mass of the scalar component of $X$ is
\begin{eqnarray}
 m_X = \sqrt{3}\frac{m_{3/2}M_{\rm Pl}}{\Lambda}.
\end{eqnarray}
This is the minimal model of the supersymmetric axion. However, the
gravity mediation potentially has problems of large CP violation which
we are trying to solve. Therefore, we consider cases with $m_{3/2} \ll
100~$GeV where the supergravity effects is subdominant.



\subsection{Model 1: spontaneous $R$-breaking model with 
$a=-1,~\Delta K =0, ~\Delta W=0$}

For $a=-1$, $X=0$ is unstable and the $R$ symmetry is spontaneously
broken by its vev. The potential is stabilized through the K\"ahler
potential $K \sim -(X^{\dag}X)^3$.  The vevs of $X$ and $A$ are
\begin{eqnarray}
\nonumber
{\langle X \rangle} &=& 
x_0 + \frac{( \tilde d - \tilde e ) (2 \tilde d- \tilde e )}{4 \tilde c }
\left( {\Lambda_c \over \Lambda} \right)^2
x_0^3 , \\
\langle A \rangle &=& \frac{2 \tilde d - \tilde e }{4 \tilde c }x_0^2.
\end{eqnarray}
At the vacuum, the masses for the CP-even scalar fields are obtained to be
\begin{eqnarray}
 m_{X_R}^2 = {2 \mu^4 \over \Lambda^2} , \ \
 m_{A_R}^2 =   {2 \tilde c \mu^4 \over f^2} 
\left( {\Lambda_c \over \Lambda} \right)^2.
\end{eqnarray}
Here $X=(x_0 + X_R )e^{iX_I}$, $A=A_R +i A_I$ and the above expressions
are correct up to of $O(x_0^4)$.  The parameter $x_0$ should satisfy a
condition\footnote{$\langle X \rangle \simeq x_0 < 1$ also satisfies the
condition to obtain the stable vacuum.  This is similar to Model 3.}
\begin{eqnarray}
 x_0 \lesssim 1,
\end{eqnarray}
for the effective theory to be valid.  Since the $R$-symmetry is
spontaneously broken, there is a nearly massless $R$-axion $X_I$.  The
$X_I$ field and the goldstino (the fermionic component of $X$) acquire
masses through a supergravity correction:
\begin{eqnarray}
m_{X_I}^2 = \frac{2\mu^4}{\sqrt{3}M_{\rm Pl} \Lambda x_0}
=2\sqrt{3}m_{3/2}^2\left(\frac{M_{\rm Pl}}{\Lambda x_0}\right)
\end{eqnarray}
up to corrections suppressed by the Planck scale and of $O(x_0^2)$.
The axion $A_I$ remains massless at this stage.

The axino $\tilde a$, which is a fermionic part of $A$, obtains a mass
via a K{\" a}hler term:
\begin{eqnarray}
 K= - \Lambda_c^2 \cdot \frac{\tilde c}{2} (A+A^{\dag})^2 X^{\dag} X.
\end{eqnarray}
It is given by
\begin{eqnarray}
 m_{\tilde a} = -
{\tilde c \mu^2 \Lambda x_0 \over f^2} 
\left( {\Lambda_c \over \Lambda }\right)^2
\end{eqnarray}
and a fermionic part of $X$, $\psi_X$, is the goldstino which is absorbed into gravitino.
Finally, the $F$ component vev for the axion is
\begin{eqnarray}
 F^A  = -
{(3 \tilde d -2 \tilde e) \over 6}\frac{ \mu ^2 \Lambda }{f^2 } 
\left( {\Lambda_c \over \Lambda }\right)^2 x_0^3 .
\end{eqnarray}
This gives the gaugino masses, $M_{1/2} = (k \alpha / 4 \pi) F^A$.


\subsection{Model 2: explicit $R$-breaking model with $a=1,~\Delta 
K \neq 0, ~\Delta W=0$}

This model is $R$-symmetric if we neglect $\epsilon_K$. Thus we have the
following scalar mass spectra up to $O(\epsilon_K^2)$:
\begin{eqnarray}
m^2_{X_R} =m^2_{X_I} =\frac{\mu^4}{\Lambda^2} ,
\end{eqnarray}
\begin{eqnarray}
m^2_{A_R} = \frac{2 \tilde c \mu^4}{f^2}
\left( {\Lambda_c \over \Lambda }\right)^2.
\end{eqnarray}
Here we denoted as $X=X_R +iX_I$.  Of course, the axion $A_I$ and
goldstino $\psi_X$ remain massless at this stage.

The shift of the vevs at the leading order in the $\epsilon_K$ expansion
are\footnote{ For small $\epsilon_K$, the SUGRA correction to $\delta X$
may have the same order of the magnitude as the following result.}
\begin{eqnarray}
\delta X &=& \epsilon_K ^3 C_3 \frac{\Lambda ^4 }{f^4}
, \\ 
\delta A &=&  \epsilon_K ^2 C_3  \frac{\Lambda^2}{2 \tilde c
 f^2} 
\left( {\Lambda_c \over \Lambda }\right)^{-2}.
\end{eqnarray}
The mass mixing between $X_R$ and $A_R$ is generated by $\epsilon_K$.
\begin{eqnarray}
 m_{XA}^2 
=  {\epsilon_K \over 2}
{\mu^4 \Lambda \over  f^3}
\left(
-6 \tilde c \left( {\Lambda_c \over \Lambda }\right)^2
+ {f^2 \over \Lambda^2}
\right).
\label{mxaia}
\end{eqnarray}
In order not for the vev shifts to be larger than $O(1)$ and also not to
destabilize the vacuum by the mass mixing, we obtain a condition
\begin{eqnarray}
 \epsilon_K < \left( {\Lambda_c \over \Lambda }\right)^2.
\end{eqnarray}


The $F$-component vev of $A$ is:
\begin{eqnarray}
 F^A 
= \epsilon_K {\mu^2 \Lambda \over f^2} .
\end{eqnarray}
This induces the gaugino masses and the axino mass through a K\"ahler
term
%
%
$({f^2 C_3}/{3!})(A+A^{\dag})^3$:
\begin{eqnarray}
m_{\tilde a} 
=
C_3 F^A  .
\end{eqnarray}


\subsection{Model 3: Explicit $R$-breaking model with $a=1,~\Delta K
  =0,~\Delta W \neq 0 $}

In this model, scalar and goldstino mass spectra are the same as Model 2
up to $O(\epsilon^2)$.  Here the vev shift of $X$ is
\begin{eqnarray}
{\delta X} = - \epsilon , \ \ \
{\delta A} = 0 .
\end{eqnarray}
Therefore, we have
a condition to make our effective theory valid:
\begin{eqnarray}
\epsilon \lesssim 1,
\end{eqnarray}
for validity of the effective theory.
Then the axino obtains a mass via $K= \Lambda_c^2\tilde c (A+A^{\dag})^2
X^{\dag}X$ as in Model~1:
\begin{eqnarray}
m_{\tilde a} 
=- \tilde c \epsilon \cdot{ \mu^2 \Lambda \over f^2} 
\left( {\Lambda_c \over \Lambda }\right)^2.
\end{eqnarray}
The $F$-component of $A$ is
\begin{eqnarray}
F^A 
=   { \epsilon^3 \tilde d \over 2} \frac{ \mu^2 \Lambda }{ f^2 }
\left( {\Lambda_c \over \Lambda }\right)^2.
\end{eqnarray}


\subsection{Summary of mass spectra and constraints on parameters in three models}

Here we summarize the mass spectra of three models in
Table~\ref{tab:mass}.  Recall we have several parameters:
\begin{eqnarray}
\nonumber
\tilde c ,~\tilde d,~ \tilde e \lesssim O(1) ,  \ \ \
C_3 = O(1).
\end{eqnarray}
There are also constraints on $R$-breaking parameters in each model:
\begin{eqnarray}
& & {\mbox{Model 1}}:  x_0 \lesssim 1 , \\
\label{cmodel1}
&& {\mbox{Model 2}}: ~
\epsilon_K  \lesssim  
\left( {\Lambda_c \over \Lambda }\right)^2,  \\
&&  {\mbox{Model 3}}: ~ \epsilon \lesssim  1 . 
\end{eqnarray}

\begin{table}[h]
\begin{center}
\newlength{\myheight}
\setlength{\myheight}{1cm}
\begin{tabular}{|c||c|c|c|c|c|c|}
\hline \parbox[c][\myheight][c]{0cm}{}
& Model 1 & Model 2 
& Model 3 
\\
\hline \parbox[c][\myheight][c]{0cm}{}
$m_{X_R}^2$ 
& $\left(\sqrt{2}F^X\right)^2$
& $\left({F^X}\right)^2$ 
& $\left({F^X}\right)^2$ 
\\
\hline \parbox[c][\myheight][c]{0cm}{}
$m_{X_I}^2$ 
& $2\sqrt{3}m_{3/2}^2\left(\frac{M_{\rm Pl}}{x_0 \Lambda}\right)$ 
& $\left(F^X\right)^2$ 
& $\left(F^X\right)^2$ 
\\
\hline \parbox[c][\myheight][c]{0cm}{}
$m_{A_R}^2$ 
& $\left(  \frac{ \Lambda_c }{f}\sqrt{2\tilde c}F^X\right)^2$ 
& $\left(  \frac{ \Lambda_c }{f}\sqrt{2\tilde c}F^X\right)^2$ 
& $\left(  \frac{ \Lambda_c }{f}\sqrt{2\tilde c}F^X\right)^2$
\\
\hline \parbox[c][\myheight][c]{0cm}{}
$m_{\tilde a}$ 
& $x_0 \tilde c   \left(\frac{\Lambda_c }{f}\right)^2 F^X  $ 
& $C_3 F_A$ 
& $ \epsilon \tilde c \left(\frac{ \Lambda_c }{f}\right)^2 F^X $ 
\\
\hline \parbox[c][\myheight][c]{0cm}{}
$F_A $ 
& $-  \frac{(2 \tilde e - 3 \tilde d  ) x_0^3 }{6}\frac{\Lambda_c^2}{f^2}F^X$ 
& $-\epsilon_K \cdot \frac{ \Lambda^2}{f^2}F^X$ 
& $- \frac{ \tilde d\epsilon^3}{2  }\frac{\Lambda_c^2}{f^2}F^X$ 
\\
\hline
\end{tabular}
\caption{A table of mass spectra in three models for a gauge mediation with light gravitino.
Here $\Lambda F^X \simeq -\mu^2 \simeq -\sqrt{3}m_{3/2}M_{\rm Pl}$ and $\Lambda_c = {\rm min}[\Lambda, f]$.
The soft mass in the visible sector are given by $M_{1/2} \simeq k
\frac{\alpha}{4\pi}F^A$ and $m_{0} \simeq \sqrt{k}
\frac{\alpha}{4\pi}F^A$.  \label{tab:mass}}
\end{center}
\end{table}

The masses in the table should be smaller than the cut-off scales,
$\Lambda / 4\pi$ in the SUSY breaking sector or $f$ in the PQ breaking
sector. Therefore, we have a constraint:
\begin{eqnarray}
{\Lambda}  > \sqrt{4\pi} \mu.
\label{call}
\end{eqnarray}


\section{Cosmological constraints}

In this section, we discuss cosmological constraints on three models
defined above.
We consider decays of the coherent oscillations of the saxion
and $X_R$ (the Polonyi field) and discuss constraints from the BBN and the
matter energy density of the universe.
We assume that the LSP is the gravitino and the next-to-lightest
supersymmetric particle (NLSP) is the bino. The bino NLSP gives more
stringent constraint than a case for stau NLSP.

In the following discussion, we fix the PQ breaking scale $f$ to be
$10^{10}$ GeV, bino mass $m_{\tilde B}$ to be $100$ GeV and all
dimensionless parameters (except for $x_0$, $\epsilon_K$ and $\epsilon$)
to be of $O(1)$.
Once we fix $f$ and the gaugino mass, the parameters we have are
$\Lambda$ and the gravitino mass (or equivalently the $R$-breaking
parameters in each model ($x_0$, $\epsilon_K$, or $\epsilon$)) aside
from the $O(1)$ parameters.  See Figure~\ref{fig1}--\ref{fig3} for the
parameter regions constrained by the validity of the effective theory.

In general, light scalar fields such as $X_R$ and $A_R$ are problematic
for cosmology since their coherent oscillations and late-time decays
would produce too large entropy and also produce unwanted particles.
In the models we are discussing, the Polonyi and the saxion can be much
heavier than the Standard Model particles due to the direct coupling to
the SUSY breaking sector. This situation helps to cure the problem. We
consider here the constraints on the model parameters from the
successfull BBN and the matter density of the
universe~\cite{Banks:2002sd, Polonyi:1977pj, Coughlan:1983ci}.


We define several important temperatures for discussion. The decay
temperature $T_d^{\phi}$ is the one at which a scalar condensate $\phi
(= X_R, X_I, A_R)$ decays in the radiation-dominated universe. The
domination temperature $T_{\rm dom}^{\phi}$ is the one at which $\phi$
dominates over the energy density of the universe \cite{Ibe:2006rc,
Heckman:2008jy, Hamaguchi:2009hy}:
\begin{eqnarray}
T_d^{\phi} &\equiv & 
\left(\frac{90}{\pi ^2 g_* (T^{\phi}_d)}\right)^{1/4}  
\sqrt{\Gamma_{\phi} M_{\rm Pl}} , \\
T_{\rm dom}^{\phi} &\equiv & 
{\rm min}[T_R,~T_{\rm osc}^{\phi}] 
\left(\frac{\Delta \phi}{\sqrt{3}M_{\rm Pl}}\right)^2 .
\label{tdom}
\end{eqnarray}
Here $g_*(T_d^{\phi})$ is the effective number of light particles at
$T_d^{\phi}$, $\Gamma_{\phi}$ is total decay width of $\phi$.  $T_R$ is
a reheating temperature by an inflaton decay, $T_{\rm osc}^{\phi} \simeq
0.3 \sqrt{m_{\phi}M_{\rm Pl}}$ is a temperature when $\phi$ starts to
oscillate and $\Delta \phi $ is the initial amplitude. These will be
important for our discussion below since $\phi$ does not dominate over
the universe if $T_d^{\phi} > T_{\rm dom}^{\phi}$.


\subsection{Model 1}
\label{ccmodel1}

We first calculate the decay widths of particles in Model 1.
\begin{itemize}
\item{$X_R$ (Polonyi) and $A_R$ (saxion) decay }
\end{itemize}
Through the interactions 
$K= \Lambda^2 [ X^{\dag}X - {(X^{\dag}X)^2}/{4} ]$,
the Polonyi field decays into $R$-axions or gravitinos.
The total decay width is given by
\begin{eqnarray}
\Gamma_{X_R} & \simeq &  \Gamma(X_R \to X_I X_I ) 
+ \Gamma (X_R \to \psi_{3/2}\psi_{3/2}) \\
& \simeq & \frac{1}{32\pi}\frac{m_{X_R}^3}{x_0^2 \Lambda^2} 
+ \frac{1}{96\pi}\frac{m_{X_R}^5}{m_{3/2}^2M_{\rm Pl}^2} .
\end{eqnarray}
The vev $x_0$ can be expressed in terms of the $R$-axion mass as
\begin{eqnarray}
 x_0 \Lambda = 2\sqrt{3}M_{\rm Pl}\left(\frac{m_{3/2}}{m_{X_I}}\right)^2.
\end{eqnarray}
%

For the saxion, the main decay modes are $A_R \to a a$ and $A_R \to
\psi_{3/2} + \tilde a$, which originate from $K =
{f^2} ({C_3}/{3}) (A+A^{\dag})^3 
- \Lambda_c^2 (\tilde c/2) (A+A^{\dag})^2 X^{\dag}X$. The decay width is
\begin{eqnarray}
\nonumber
\Gamma_{A_R} &\simeq & \Gamma (A_R \to a a ) + \Gamma (A_R \to  \psi_{3/2} + \tilde a )\\
& \simeq & \frac{C_3^2}{32 \pi} \frac{m_{A_R}^3}{f^2} + \frac{1}{96 \pi} \frac{m_{A_R}^5}{m_{3/2}^2M_{\rm Pl}^2} .
\end{eqnarray}
The axino ($\tilde a$) produced by the saxion decay subsequently decays into
gravitinos or the Standard Model particles with the decay width:
\begin{eqnarray}
\Gamma_{\tilde a} &\simeq & \Gamma(\tilde a \to \psi_{3/2} + a)
+ \Gamma(\tilde a \to \lambda + g)  \\ 
&\simeq &
\frac{1}{96\pi}\frac{m_{\tilde a}^5}{m_{3/2}^2M_{\rm Pl}^2} 
+ N_g \frac{\alpha^2}{256 \pi^3} \frac{m_{\tilde a}^3}{f_a^2} ,
\end{eqnarray}
where $N_g$ (=12) is the number of the decay modes\footnote{ There is
also a decay mode $\tilde a \to \psi_{3/2} + X_I$, which can be the main
decay mode in a narrow region in the parameter space.}.


Just by looking at the main decay modes, one can see that it is
problematic if the coherent oscillations of the Polonyi field or the
saxion field dominate over the energy density of the
universe\footnote{If the saxion $s = f e^{A}$ is captured at the origin
$s=0$ during the inflation, the thermal effect via messenger
fields becomes relevant to the saxion potential.  For such a case, the
saxion can dominate the energy density of the
universe~\cite{Nakamura:2008ey}.}.  The decay products are stable (or
long-lived) particles such as gravitinos and axions which contributes to
the matter energy density of the universe. The overproduction of those
particles needs to be avoided for viable cosmology.

For of $O(M_{\rm Pl})$ initial amplitudes of the Polonyi and the saxion,
the condition for the matter density $\Omega_{\rm matter} \sim 0.2$
requires a low enough reheating temperature after inflation such as $T_R \lesssim 1$
MeV. Such a low reheating temperature may be barely consistent with the BBN.
However, we do not consider this case 
since $O(M_{\rm Pl})$ field values are beyond the validity of the effective theory.
%
%
%
One can consider possibilities that the initial amplitudes are at
$\Delta X_R =O(\Lambda)$ \cite{Ibe:2006am} or $\Delta A_R =O(f)$
\cite{Carpenter:2009sw}, since $\Lambda$ and $f$ are the unique
(cut-off) scales for $X$ and $A_R$, respectively\footnote{A possiblity
$\Delta A_R = O( \sqrt{f_a M_{\rm GUT}})$ was also considered in
Ref.~\cite{Hasenkamp:2010if}.  }.
For such a case, the Polonyi (the saxion) decays before dominating the
universe, with the parameter region that we are considering, since
conditions $T_{\rm dom}^X < T_d^X$ and $T_{\rm dom}^{A_R} < T_d^{A_R}$
are always satisfied. 
Hereafter we consider such a case to search for viable parameter regions.

%
%


\begin{itemize}
\item $X_I$ ($R$-axion) decays
\end{itemize}
The $R$-axion decays into two gauginos if it is kinematically allowed. 
The interaction term in the Lagrangian is~\cite{Hamaguchi:2009hy}
%
\begin{eqnarray}
-i \frac{X_I}{\sqrt{2} x_0 \Lambda} \frac{M_{1/2}}{2 }\lambda \lambda  +c.c. .
\end{eqnarray}
Here $\lambda$ is the MSSM gauginos.
The total decay width is
\begin{eqnarray}
\Gamma_{X_I}  &\simeq & \Gamma (X_I \to \lambda \lambda) \\
&\simeq &
\frac{N_g}{32\pi} m_{X_I} \left(\frac{M_{1/2}}{x_0 \Lambda}\right)^2 \left( 1- \frac{4M_{1/2}^2}{m_{X_I}^2}\right)^{1/2} .
\end{eqnarray}

For $m_{X_I} < 2M_{1/2}$, the channel $X_I \to \lambda \lambda$ is
closed, then $X_I \to b \bar{b}$ is the main decay mode 
through the mixing between $X_I$ and the CP-odd Higgs boson ($A$) in the MSSM\footnote{
We neglected a decay mode $X_I \to t \bar{t}$ via the similar interaction, since we assumed $m_t > m_{\tilde B} = 100$ GeV.
On the other hand,
note also that so long as $\tan \beta > \sqrt{m_t/m_b} \approx 6.4$ this mode is suppressed, compared to $X_I \to b \bar{b}$.
In a whole computation, we assumed $m_A$ and $m_{X_I}$ do not degenerate.
}. 
The mixing is obtained through the $B\mu$-term.  Although we need a concrete
model to generate $B\mu$-term to discuss that interaction, there is
always a contribution from a one-loop diagram with the gaugino mass
insertion~\cite{Martin:1997ns, Babu:1996jf}, even if we have $B\mu
=0$ at the messenger scale~\cite{Dine:1996xk}.
If that is the donimant contribution, there is an interaction term to
mix $X_I$ and $A$:
\begin{eqnarray}
\frac{i}{2\sqrt{2}}\frac{m_A^2 \sin 2 \beta}{x_0 \Lambda} X_I H_u H_d +c.c.,
\end{eqnarray} 
where we have used a relation $B \mu = m_A^2 \sin 2 \beta /2$.
From this interaction and Yukawa coupling, the decay width is found
to be~\cite{Hamaguchi:2009hy}:
\begin{eqnarray}
\Gamma_{X_I} \simeq \Gamma (X_I \to b \bar b) 
&\simeq & 
\frac{3m_{X_I}}{16 \pi}\left(
\frac{m_A^2 \sin^2 \beta}{x_0 \Lambda} \frac{m_b}{m_A^2-m^2_{X_I}}
\right)^2 \sqrt{1-{4m_{b}^2 \over m_{X_I}^2}}.
\end{eqnarray} 
For further smaller $m_{X_I}$, 
$X_I \to \tau \bar \tau$ can be the main decay mode with large $\tan
\beta$; $m_b$ should be replaced with $m_{\tau}$.
In the parameter region that we are interested, $T_d^{X_I} \gtrsim 1$
MeV\footnote{
If the decay of $X_I$ is too late, it can influence the BBN. This excludes 
a small parameter region around $\Lambda \sim 10^{13}$~GeV and $m_{3/2} \sim 40$~MeV 
in Figure~1~\cite{Kawasaki:2004yh}.
}.

On the other hand, because $x_0 \Lambda$ is a normalization of the $R$-axion,
its initial amplitude is at most on the order of $x_0 \Lambda$.  
%
The temperature at which oscillating $R$-axion dominates the universe is
given by
\begin{eqnarray}
T_{\rm dom}^{X_I} 
\nonumber
&=& \frac{1}{4}{\rm min}[T_R,~T_{\rm osc}^{X_I}]
\left(\frac{m_{3/2}}{m_{X_I}}\right)^4 
\left(\frac{\Delta X_I}{x_0
 \Lambda }\right)^2 \\
\nonumber
& \lesssim & 0.16 {\rm MeV} 
\left(\frac{{\rm min}[T_R,~T_{\rm osc}^{X_I}]}{10^{10}{\rm GeV}} \right)
\left(\frac{m_{3/2}/m_{X_I}}{5\times 10^{-4}}\right)^4
\left(\frac{\Delta X_{I}}{x_0 \Lambda}\right)^2 .
\end{eqnarray} 
Here $T_{\rm osc}^{X_I} \simeq 0.3 \sqrt{m_{X_I}M_{\rm Pl}} \simeq 1.6
\times 10^{10} {\rm GeV} \left({m_{X_I}}/{1.2{\rm
TeV}}\right)^{1/2}$ is a temperature where the $R$-axion starts to
oscillate.  
The relation $m_{3/2}/m_{X_I} \lesssim 5\times 10^{-4}$ holds for
$m_{3/2} \lesssim 40$~MeV which we need from the BBN constraint we
discuss later.
Then we obtain $T^{X_I}_d \gg T_{\rm dom}^{X_I}$ in our model; the
$R$-axion does not dominate over the universe.


\begin{itemize}
\item{Bino NLSP and BBN constraint}
\end{itemize}
We have seen that the Polonyi field, the saxion, or $R$-axion do not
dominate over the energy density of the universe provided that the
initial amplitudes of the Polonyi field and the saxion are of
$O(\Lambda)$ and of $O(f)$, respectively.  Then we need to consider the
bino abundance since it is a long-lived NLSP, which can disturb the BBN.
%
%
The thermal abundance of binos is given by~\cite{Feng:2004zu}
\begin{eqnarray}
Y_{\tilde B}^{\rm th} = \frac{n_{\tilde B}}{s} =
4\times 10^{-12} \times \left(\frac{m_{\tilde B}}{100 {\rm GeV}} \right),
\label{ybino}
\end{eqnarray} 
below the freeze-out temperature $T_f^{\tilde B} \sim m_{\tilde B}/30$.

\begin{figure}[htbp]
 \begin{center}
  \includegraphics[width=100mm]{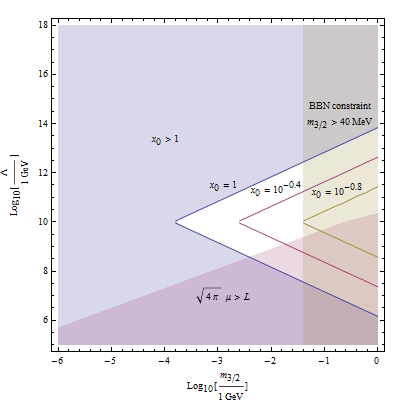}
 \end{center}
 \caption{A figure of allowed parameter region for $M_{1/2} = 100$ GeV.
Recall that we have constraint that should satisfy $\sqrt{4\pi} \mu
\lesssim L$, $x_0 \lesssim 1$ and $m_{3/2} \lesssim 40$ MeV. Here $L =
(\sqrt{2}x_0)^{1/2}\Lambda \lesssim \Lambda$.}  \label{fig1}
\end{figure}

Since the binos decay into gravitino at a later time, there is
constraint from the BBN on the lifetime as mentioned above. With the
yield of Eq.~(\ref{ybino}), this can be translated to the bound on the
gravitino mass which is~\cite{Kawasaki:2008qe}
\begin{eqnarray}
m_{3/2} \lesssim 40~{\rm MeV} \ \ \ {\rm for} \ m_{\tilde B} =100 ~{\rm GeV}.
\end{eqnarray} 
For a larger bino mass or for the stau NLSP case, the constraint is
relaxed, such as $m_{3/2} \lesssim 1$ GeV.
See Figure~\ref{fig1} for the allowed region.

There are contributions to the bino density from the decays of the heavy
field such as $X_R$, $A_R$ and $\tilde a$.
Since they are heavy enough, the binos produced by those decays are
thermalized and thus such contributions are already taken into account
in Eq.~(\ref{ybino}). The decays of $X_I$ can be later than the
freezing-out of the bino-pair annihilation. This non-thermal
contribution is smaller than the thermal piece when
\begin{eqnarray}
 T_R \lesssim 10^6~{\rm GeV},
\end{eqnarray}
which is required later by the constraint from the gravitino abundance.
%
%

In summary in Model 1 the gravitino mass is constrained to be 
\begin{eqnarray}
 10^{-4}~{\rm GeV} \lesssim m_{3/2} \lesssim 40~{\rm MeV},
\label{eq:grav-range}
\end{eqnarray}
and the cut-off scale (or dynamical scale) of the SUSY breaking sector
is
\begin{eqnarray}
 10^8~{\rm GeV} \lesssim \Lambda \lesssim 10^{13}~{\rm GeV},
\label{eq:lam-range}
\end{eqnarray}
for $f=10^{10}$~GeV.


\begin{itemize}
\item{Candidates of cold dark matter}
\end{itemize}

The thermal contribution to the gravitino density
$\Omega_{3/2}^{\rm th}$ is
\cite{Moroi:1993mb}:
\begin{eqnarray}
\Omega^{\rm th}_{3/2}
& = &  0.2 \left(\frac{T_{R}}{3.1 \times 10^{6} {\rm GeV}} \right)
\left(\frac{30 {\rm MeV}}{m_{3/2}} \right)
\left(\frac{M_{\tilde g}}{600 {\rm GeV}} \right)^2 .
\label{omega32}
\end{eqnarray} 
Here $M_{\tilde g}$ is gluino mass at TeV scale.  Once one fixes gluino
mass, the gravitino by thermal scattering can be dark matter of the
universe for $T_R \lesssim O(10^{6})~{\rm GeV} \times \left(
{m_{3/2}}/{O(10)~{\rm MeV}}\right)$.

Note that we also have non-thermal component to the gravitino density.
The important contribution is from the Polonyi decay:
\begin{eqnarray}
Y_{3/2}^{X_R} \simeq \frac{3 T_{\rm dom}^{X_R}}{2 m_{X_R}}B^{X_R}_{3/2} .
\end{eqnarray}
Here $B^{X_R}_{3/2}$ is a branching ratio of a decay mode $X_R \to 2\psi_{3/2}$.
Then we obtain
%
%
%
\begin{eqnarray}
\Omega_{3/2}^{\rm NT}
 \simeq  0.23 \left(\frac{\Lambda}{10^{13}{\rm GeV}}\right)^3
\left(\frac{T_R}{10^6{\rm GeV}}\right)B_{3/2}^{X_R} .
\end{eqnarray} 
Here $B_{3/2}^{X_R} = O(1)$.  For $\Lambda \simeq 10^{13}$~GeV and
$m_{3/2} \simeq 40$ MeV (near the upper right boundary of the
Figure~\ref{fig1}), the non-thermal component can be as important as the
thermal one.
%
%
Considering the BBN constraints, the contribution to the gravitino
density from the decays of binos, saxions, and the axinos\footnote{
Recently, a paper \cite{Cheung:2011mg} discussed the constraint on the reheating temperature 
obtained from the matter energy density of the thermally produced axinos \cite{Strumia:2010aa} 
or gravitinos from their decays.
In our model, the gravitino abundance from the thermally produced axino is given by
\begin{eqnarray}
\nonumber
\Omega_{3/2}^{\tilde a} \simeq
0.2 \left(\frac{T_R}{8.7\times 10^5 {\rm GeV}}\right)
\left(\frac{m_{3/2}}{30 {\rm MeV}}\right)
\left(\frac{ 10^{12} {\rm GeV}}{f_a}\right)^2 B_{3/2}^{\tilde a} .
\end{eqnarray}
Here we have replaced $M_{\tilde g}$ with $\sqrt{6}\alpha_3 m_{3/2}M_{\rm Pl}/(4\pi f_a)$ 
and used $g_3 \simeq 1$ in eq.(\ref{omega32}), and $B_{3/2}^{\tilde a}$ is a branching ratio of axino decay to gravitino.
As 
one always finds $B^{\tilde a}_{3/2}/f_a^2 \gtrsim 10^{-24}~ {\rm GeV}^{-2}$ for $f_a \lesssim 10^{12}$ GeV,
this can give more stringent constraint on $T_R$ together with eq.(\ref{omega32}).
The allowed region for the reheating temperature is
\begin{eqnarray}
\nonumber
T_R < m_{\tilde a}, ~~~{\rm or} ~~~
m_{\tilde a} < T_R < 8.1\times 10^5 {\rm GeV} 
\left(\frac{f_a}{10^{12}{\rm GeV}} \right)\left(\frac{600{\rm GeV}}{M_{\tilde g}}\right)
\frac{1}{\sqrt{B^{\tilde a}_{3/2}}} .
\label{axinocond}
\end{eqnarray}
} 
is much smaller than this contribution for $T_R \lesssim O(10^6)$ GeV and
$m_{3/2} \lesssim 40$ MeV.  This is because of low $T_{\rm dom}$ and
small branching ratios into gravitinos.

The axion is also a candidate for dark matter~\cite{Kim:1986ax}.
The abundance is given by
\begin{eqnarray}
\Omega_a \simeq 1.4 \left(\frac{\Theta_{\rm mis}}{\pi} \right)^2 \left(\frac{f_a}{10^{12}{\rm GeV}} \right)^{7/6} ,
\end{eqnarray} 
where $\Theta_{\rm mis}$ is misalignment angle of the axion.


\subsection{Model 2 and Model 3}
\label{ccmodel23}

Essentially, the discussion is parallel to Model 1. A difference is that
the $R$-axion now has a similar mass to the Polonyi field, and thus we
do not need to consider it separately
As in Model 1, the domination of
the Polonyi and the saxion fields would produce too much gravitinos, and
therefore the initial amplitude should be small enough. With
$O(\Lambda)$ and $O(f)$ for the sizes of the initial amplitudes, the
problem can be avoided.

\begin{figure}[htbp]
 \begin{center}
   \includegraphics[width=100mm]{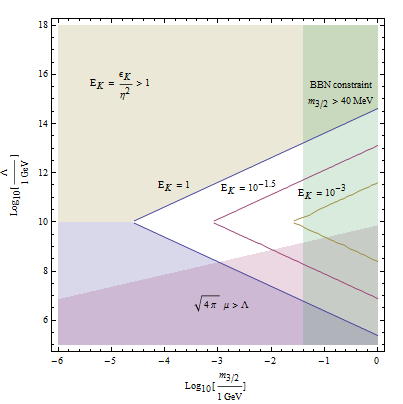}
 \end{center}
 \caption{A figure for an allowed parameter region for Model 2.
This is similar to the Model 1 except for $\epsilon_K \lesssim \eta^{2}.$ 
Here we defined $\eta = \frac{\Lambda_c}{\Lambda}$ and $E_K = \epsilon_K \eta^{-2}$.}
 \label{fig2}
\end{figure}

\begin{figure}[htbp]
 \begin{center}
   \includegraphics[width=100mm]{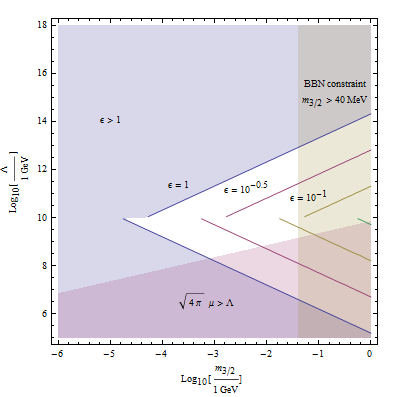}
 \end{center}
 \caption{The similar figure for Model 3.  This is similar to the
Model 1 and 2 except for $\epsilon \lesssim 1.$ The discontinuity of
$\epsilon$ originates from the fact $F^A \propto \tilde{d} = (d_1
+2q_X)$ and we took $q_X =1$ for $f>\Lambda$ while $q_X=0$ for $f <
\Lambda$.  } \label{fig3}
\end{figure}

The viable parameter regions are shown in Figures~\ref{fig2} and
\ref{fig3} which are very similar to Figure~\ref{fig1}. Therefore, the
viable range we obtain is the same as Eqs.~(\ref{eq:grav-range}) and
(\ref{eq:lam-range}).
We summarize in Table~\ref{table2} the numerical values of the masses
and decay widths in the parameter range of our interest.

%
%

\begin{table}[h]
\begin{center}
\setlength{\myheight}{1cm}
\begin{tabular}{|c||c|c|c|c|c|c|}
\hline \parbox[c][\myheight][c]{0cm}{}
& Model 1 & Model 2 
& Model 3 
\\
\hline \parbox[c][\myheight][c]{0cm}{}
$m_{X_R}$ (GeV)
& $10^5 - 10^8$
& $10^4 - 10^8$ 
& $10^4 - 10^8$ 
\\
\hline \parbox[c][\myheight][c]{0cm}{}
$\Gamma_{X_R} 
\simeq \Gamma_{X_R \to 2\psi_{3/2}}~(+\Gamma^{\rm M1}_{X_R \to 2X_I})$ (GeV)
& $10^{-10} - 10^4$
& $10^{-16} - 10^4$ 
& $10^{-16} - 10^4$ 
\\
\hline \parbox[c][\myheight][c]{0cm}{}
$m_{X_I}$ (GeV)
& $5 - 5\times 10^{3}$ 
& $10^4 - 10^8$ 
& $10^4 - 10^8$ 
\\
\hline \parbox[c][\myheight][c]{0cm}{}
$\Gamma_{X_I}
\simeq  \Gamma^{\rm M 1}_{X_I \to 2\lambda}~{\rm or}~\Gamma^{\rm M 1}_{X_I \to 2b}$ or
$\Gamma^{\rm M 2,3}_{X_I \to 2\psi_{3/2}}$ (GeV) 
& $10^{-24} - 10^{-11}$
& $10^{-16} - 10^4$ 
& $10^{-16} - 10^4$ 
\\
\hline \parbox[c][\myheight][c]{0cm}{}
$m_{A_R}$ (GeV)
& $10^5 - 10^7$ 
& $10^4 - 10^8$  
& $10^4 - 10^8$ 
\\
\hline \parbox[c][\myheight][c]{0cm}{}
$\Gamma_{A_R}
\simeq \Gamma_{A_R \to 2a}+\Gamma_{A_R \to  \psi_{3/2}\tilde a}$ (GeV)
& $10^{-7} - 10$
& $10^{-10} - 10$ 
& $10^{-10} - 10$ 
\\
\hline \parbox[c][\myheight][c]{0cm}{}
$m_{\tilde a}$ (GeV)
& $5\times 10^4 - 10^6$ 
& $10^4$ 
& $10^4 - 10^6$ 
\\
\hline \parbox[c][\myheight][c]{0cm}{}
$\Gamma_{\tilde a}
\simeq \Gamma_{\tilde a \to \psi_{3/2}a}+\Gamma_{\tilde a \to \lambda g}$ (GeV)  
& $10^{-12} - 10^{-5}$
& $10^{-12} - 10^{-10}$ 
& $10^{-12} - 10^{-6}$
\\
\hline \parbox[c][\myheight][c]{0cm}{}
$R$-breaking parameter: $(x_0,{\epsilon_K \over \eta^{2}},\epsilon)$ 
& $10^{-0.8}-1$ 
& $10^{-3}-1$ 
& $10^{-1}-1$ 
\\
\hline
\end{tabular}
\caption{A table of examples of numerical values in each model. We took
all dimensionless parameters as of order unity, $M_{1/2} = 100$ GeV,
$f=10^{10}$ GeV and $m_{3/2} \lesssim 40$ MeV. 
Here we defined $\eta = \Lambda_c /\Lambda$.} \label{table2}
\end{center}
\end{table}


\section{Brief comments on axions in IIB orientifold/F-theory GUTs}

In string therory, besides the field theoretic axions, 
we often obtain very light string theoretic axions via moduli stabilization in the low scale
SUGRA \cite{Arvanitaki:2009fg, Abe:2005pi}. In general, the number of axions is estimated as
\begin{eqnarray}
\nonumber
{\mbox{(The number of axions)}} = {\mbox{(The number of moduli fields) $+1 -$ (The number of terms in the $W$)}}.
\end{eqnarray}
Here $W$ is the superpotential and a factor unity comes from the
$R$-symmetry. This is because PQ symmetries of (moduli) fields and
the $R$-symmetry produce candidates of the axion whereas independent terms in the superpotential
kill them, supposing the K\"ahler potential $K$ preserves these
symmetries. When this counting gets negative or zero, we do not have any light axions.
If there are very small subleading terms violating PQ
symmetries in $W$ or $K$, they give very light mass to the axions.
If we are to identify one of the axions with the QCD axion, 
the quality of the PQ symmetry needs to be checked for solving the strong CP problem:
$\delta m_{a}^2 \lesssim 10^{-11}(m_{a}^{\rm QCD})^2$. 
Here axion mass $\delta m_{a}^2 $ is a contribution from non-QCD effects
and $(m_{a}^{\rm QCD})^2$ is the QCD axion mass just from the instanton.

For a string theoretic (QCD) axion, we often encounter a logarithm type
K\"ahler potential like $K_{\rm axion} = -M_{\rm Pl}^2\log(A+A)$, 
while we can see also quadratic K\"ahler potential $K_{\rm axion}
=\frac{f_s^2}{2}(A+A^{\dag})^2$ near the singularity.  
For logarithm case,
by the expansion around the vev like the standard form, we can obtain
the quadratic one with a decay constant 
$f = \frac{M_{\rm Pl}}{(\langle
A+A^{\dag}\rangle )} \sim f_s = O(M_{\rm string})$, where $M_{\rm string}$ is the string scale.
Then we can also obtain $K = (X^{\dag}X)^2/f^2$
after integrating out a massive gauge boson of (non-)anomalous $U(1)_{\rm PQ}$\footnote{
For non-anomalous $U(1)$ the gauge boson mass is lower than the string scale bacause a matter-like field
always cancels its Fayet-Iliopoulos term; in general we obtain $K = (X^{\dag}X)^2/m_{U(1)}^2$, 
where $m_{U(1)}$ is the (non-)anomalous $U(1)$ gauge boson mass.
}.
Its contribution to the $X$ mass is on the order of $F^X/f$. 
This corresponds to the case with $\Lambda \sim f$, once we could manage to obtain $f$ at an intermediate scale.
One should also consider the cosmological constraints since there typically exist light moduli in a low scale SUSY breaking scenario
\cite{Conlon:2007gk}.
(See also a recent paper \cite{Dine:2010cr}.)

In Ref.~\cite{Heckman:2008qt}, it has been studied a model which has the same superpotential as in Eq.~(\ref{eq:super}).
There, $U(1)_{\rm PQ}$ is an anomalous gauge symmetry at a high energy scale,
and $X$ is the Polonyi field.
$A$ is a string theoretic axion multiplet and is describing the 4-cycle
volume on which a D7-brane holding the $U(1)_{\rm PQ}$ is wrapping.  
In the paper $\langle X \rangle = f_a/\sqrt{2}$ is obtained and $A$ is
absorbed into the $U(1)_{\rm PQ}$ gauge multiplet via $D$-term moduli stabilization
since we have $f_a \ll f_A$.
Here $f_A$ is a decay constant of $A$. 
Hence, at low energy, the relevant field is only the Polonyi field,
$X$; the field ${\rm Arg}(X)$, the $R$-axion, is identified as the QCD
axion.
However, from the discussion in Refs.~\cite{Carpenter:2009sw} and
\cite{Dine:2009sw}, the axion becomes either massive or the
decay constant is unacceptably large.
Therefore we need to modify the superpotential to have the QCD axion
successfully.

For this purpose, we may consider an effective superpotential on intersecting seven branes induced by stringy instantons
\cite{Witten:1996bn}, for instance:
\begin{eqnarray}
W 
&=& \epsilon M_{\rm Pl} s  X  e^{-q_{B} B} 
+ X' (s\cdot s' - \epsilon' M_{\rm Pl}^2 e^{-q_{B}B}) + s' \Psi \bar\Psi + \cdots
\end{eqnarray}
with $U(1)_{\rm PQ} \times U(1)_{\rm PQ'}$ symmetry where both $U(1)$ symmetries are anomalous.
Here $\Psi$ and $\bar\Psi$ are messenger fields,
dots represent constant $W_0$, heavy fields in the SUSY breaking sector or in the PQ breaking sector, 
the moduli stabilization sector and so on. $X'$ is a Lagrange multiplier superfield.
$\epsilon$ and $\epsilon '$ are complex structure moduli/dilaton and other moduli contributions\footnote{
We will assume $\epsilon$ and $\epsilon '$ are singlet under two $U(1)$s for simplicity below, 
and they might become important to get proper scales.
On the other hand, instead of $\epsilon ' e^{-q_{B}B}$,
we could have an another moduli contribution $\epsilon '' e^{-q_{B'} B'}~(B' \neq B)$ for instance.
} besides $B$.
$B$ is now a string theoretic axion multiplet
describing a proper 4-cycle which would not intersect with a D7-brane holding $U(1)_{\rm PQ}$ 
(or its orientifold image brane),
and transforms non-linearly under $U(1)_{\rm PQ'}$;
we may find 
$PQ(e^{-q_{B} B})=PQ(X')=0$, $ PQ(X) = q_X = -q_{\Psi \bar\Psi}$ and
$q_B = PQ'(s) + PQ'(X) = PQ'(X') \neq 0$.
$s$ and $s'$ may appear as
$s = f_A e^{- q_X A}$ and $s' = f_A e^{ q_{X} A}$ through a constraint
where we defined $ M_{\rm Pl}^2 \langle \epsilon'  e^{-q_{B}B}\rangle \equiv f_A^2 = O(f_a^2)$
and we obtain $X'=0$ via an $F$-term equation.
Stabilization of these fields, however, can depend on the model.
For instance, let $\xi_{U(1)_{\rm PQ}}$ and $\xi_{U(1)_{\rm PQ'}}$ be Fayet-Iliopoulos (FI) terms in $U(1)_{\rm PQ}$ and
$U(1)_{\rm PQ'}$ respectively.
Note $\xi_{U(1)_{\rm PQ'}}$ includes $B$ while $\xi_{U(1)_{\rm PQ}}$ can depend on other modulus, say $C$.
Then $D$-term potential (with non-trivial world volume flux)
would almost fix $C$ and $B$ at $\xi_{U(1)_{\rm PQ}} \simeq 0$ and $\xi_{U(1)_{\rm PQ'}} \simeq 0$,
leaving $X$ and $A$ light modes\footnote{
We assumed an instanton depending on $C$ does not have two fermionic zero modes related with 
${\cal N}=1$ supercharge broken by the instanton
or its coefficient in the superpotential has a vanishing vev;
$\partial_C K =0$ could mean the potential minimum
and $\partial_B K W + M_{\rm Pl}^2 \partial_B W =0$ would be simultaneously satisfied there, though
$\xi \lesssim f_A^2$ would be sufficient for us to get a proper axion decay constant.
}. 
Thus each $U(1)$ gauge multiplet would get massive by eating $B$ and $C$, assuming
decay constants of $B$ and $C$ are of $O(M_{\rm string})$ which are larger than $f_A$ and $|\langle X \rangle|$.
Hence we would have the global $U(1)_{\rm PQ}$ symetries in the low energy scale\footnote{
We could have another possibilities without using an anomalous $U(1)$ gauge symmetry.
For accidental axion models in heterotic orbifold,
see \cite{Choi:2009jt}. 
For a discrete $R$-symmetry argument which heterotic string models might have, 
see \cite{Lee:2011dy}.},
though another global $U(1)_{\rm PQ'}$ will be broken by a stringy instanton.
Then we could gain $W = W_0 + \mu^2 Xe^{- q_X A}$ in the low energy.
Further analyses would be beyond our scope since they depend on string model
including issues of string moduli stabilization (see for instance \cite{Choi:2010gm}). 

Finally, we would like to mention moduli stabilization in type IIB
orientifold models.  
%
For moduli mediation scenario, we can get (more than) TeV scale moduli masses.
This is because moduli masses are typically related
with SUSY breaking gravitino mass because of axionic shift symmetries.
They should be heavy enough to avoid cosmological problems.
For gauge mediation scenario, we may need light gravitino mass and may get subsequently light moduli.
Hence, for such a case with a field theoretic QCD axion,
supersymmetric moduli stabilization which breaks the shift symmetry should be considered, such that
we obtain heavy moduli:
$\langle W \rangle\approx 0,~\langle \partial_{\rm moduli}W \rangle
\approx 0$ via fine-tuning \cite{Kallosh:2004yh}. For instance, $W=W_0
+ \sum_i^{h_{1,1}^{+}} (A_ie^{-a_i T_i} +B_ie^{-b_i T_i}) $ with
a fine-tuning of fluxes $W_0 < 1$ or non-geometric flux compactifications
\cite{Shelton:2005cf} might be viable. To obtain large volume, $T_i \gg
1$, we will need further fine-tunings among $A_i,~B_i,a_i,~{\rm and}~b_i$. 
The remaining K\"ahler moduli may be stabilized via $D$-terms as mentioned above.
Then we get $F^{\rm moduli} =0$ and much smaller constant term in the superpotential to obtain light gravitino mass, 
and moduli mass are decoupled to the gravitino mass: $\langle W \rangle \ll \langle
\partial^2_{\rm moduli} W \rangle$ in the Planck unit.


\section{Discussion and conclusion}

We discussed supersymmetric effective theories of the axion field and
the Polonyi field, which can be charged under global $U(1)_{\rm PQ}$
symmetry.  
We assumed that 
the SUSY breaking and the PQ breaking sectors are directly coupled 
whereas the visible sector is communicate with those sectors through messenger fields.
For a concreteness, we construct
theories by Taylor expansion in $X$ and $A$.  To compute those in a
simpler way or as generalizations, we transformed theories to ``the
standard form'' with (partial) $U(1)_{\rm PQ}$ transformation after
expansion around the saxion vev $\langle A_0 \rangle$.  Then we
considered three $R$-breaking models to obtain large gaugino mass in a
gauge mediation with light gravitino mass such like $m_{3/2} \lesssim
40$ MeV to avoid BBN constraints.
In those models, axion multiplet plays important roles in mediating
axion and SUSY breaking to the visible sector.  Both the Polonyi and
saxion can be heavy because the former has the low scale cut-off
$\Lambda$ and the latter has a direct coupling to the Polonyi
field. However, they must not dominate the universe after the inflation.
%
%
%
%
If the initial amplitude is of the order of the cut-off scale of the
effective theory, such dominations can be avoided.
%
%
For $T_R \lesssim O(10^{6})$ GeV or $T_R \lesssim m_{\tilde a}$, the axion or gravitino can be dark
matter of the universe.

We have several things we did not consider here explicitly.  For
instance, since inflaton can decay to gravitinos~\cite{Endo:2007sz}, the
resulting gravitino abundance may give an affect on our study.  But this
can depend on the inflation model.  We will also need to discuss the
generation of $\mu$-term and $B\mu$-term, which is related to the main
decay mode of the $R$-axion and other fields.  This issue is common in
gauge mediation models. 
(See recent gauge mediation models with the QCD axion \cite{Choi:2011rs},
in which messenger fields of SUSY breaking and axion are not unified. 
With $W_{\mu} = \mu' e^{-q A}H_u H_d$ \cite{Kim:1983dt} and with $F^A \simeq 0$ up to $m_{3/2}$,
$\mu/B\mu$ problem have been solved there.)

As future directions, we can also study another parameter region, where
some dimensionless parameters are much smaller than of order unity.  
For such a case, new possibilities may open up; for instance, we
may have much smaller masses of saxion and axino than our cases. 
The saxion dominated universe may be allowed in such cases because of the different decay properties.
It will be also interesting to study the UV completion, 
where the SUSY breaking sector and the PQ breaking sector are
unified.
That will be a minimal model to solve hierarchy problem and
strong CP problem.  

{\bf Acknowledgement}

We would like to thank Joe Marsano for fruitful discussions at the early
stage of this work.
We would like to thank K.S. Choi, M. Cicoli and
T. Kimura for helpful discussions about F-theory GUTs, LARGE volume
scenario and flux compactifications. 
We would like also to thank Wilfried Buchm\"uller for reading this manuscript.
RK would like to thank the DESY
theory group for the hospitality during his stay. 
He would also like to thank the Yukawa Institute for Theoretical Physics for supporting his stay in March, 2011.
RK is supported in part by the Grant-in-Aid for Scientific Research 21840006 of JSPS.


\appendix

{\Large \bf Appendix}

\section{Supergravity effects}
\label{SUGRA}

In SUGRA, $R$-symmetry can be broken by a constant term in the superpotential, 
$\delta W = \mu^2 W_0$. The scalar potential is 
\begin{eqnarray}
V_{\rm SUGRA} = e^{K/M_{\rm Pl}^2}\left[K^{i\bar{j}}D_iW \overline{D_jW} -3\frac{|W|^2}{M_{\rm Pl}^2}\right],
\end{eqnarray}
where $D_i W = \partial_{i}W + W\frac{(\partial_i K)}{M_{\rm Pl}^2}$.
The gravitino mass is given by $m_{3/2} = \mu^2 W_0 / M^2_{\rm Pl}$.
Here the K\"ahler potential and the superpotential are given by eq.(\ref{standardw}) and (\ref{standardk}).
Hereafter in the appendix we will define 
\begin{eqnarray}
\eta \equiv \frac{\Lambda_c}{\Lambda} .
\end{eqnarray}

\begin{itemize}
\item $R$-symmetric case: $a=1$ 

In this case, the vev shift is
\begin{eqnarray}
\nonumber
\delta X = {2 \Lambda \over \sqrt{3} M_{\rm Pl}} 
\cdot { \sqrt{3} W_0 \over M_{\rm Pl}} , \ \
\delta A= -\frac{c_0 \left(\eta^{-2} f^2 +2 \tilde c \Lambda ^2\right)}{3 \tilde c M_{\rm Pl}^2} 
.
\end{eqnarray}
Here
$c_0$ is a coefficient of the linear term in the K\"ahler potential, $c_0 f^2 (A + A^\dagger)$, and
in order to make our perturbation valid, we need $\frac{\Lambda}{M_{\rm Pl}} \ll \frac{f}{\Lambda}$ for $f \ll \Lambda$.
The constant term $W_0$ is fixed by the condition $V_{\rm SUGRA}=0$:
\begin{eqnarray}
 W_0 \simeq {1 \over \sqrt{3}}M_{\rm Pl}.
\end{eqnarray}
Then we find $\Lambda F_X= -\mu^2 = -\sqrt{3}m_{3/2}M_{\rm Pl}$.
The axino obtains a mass
\begin{eqnarray}
 m_{\tilde a } =  m_{3/2} \left(1 + \eta^{2} 
{2 \tilde c \Lambda^2 \over f^2} \right).
\end{eqnarray}
The first $m_{3/2}$ contribution vanishes when we replace the axion kinetic
       term, $(A+A^\dagger)^2/2$,
       with $\exp(A+A^\dagger)$. 
This is because we can define $A_{\rm new} = A +{\rm log}\Phi_c$, where $\Phi_c$ is a conformal compensator.
Hence loop effects, such that $c_1 '  (A+A^{\dag})\cdot \exp(A+A^\dagger)$, are important.
The axion field $A$ gets an $F$-component vev at the leading order of $1/M_{\rm Pl}$:
\begin{eqnarray}
 F^A = c_0 m_{3/2} .
\end{eqnarray}


\item Spontaneous $R$-symmetry breaking case: $a=-1$ and $x_0 \neq 0$

There are similar vev shifts and $F^A$ to the previous case:
\begin{eqnarray}
\nonumber
\delta X = \frac{\Lambda }{\sqrt{3}M_{\rm Pl}} 
+\frac{c_0 q_X }{3 \tilde c} \left(\frac{f}{\eta M_{\rm Pl}}\right)^2 x_0 , \ \ \
\delta A= 
-\frac{c_0 (f^2 \eta^{-2}+\tilde c\Lambda^2)}{3\tilde c M_{\rm Pl}^2 }
-\frac{(q_X\eta^{-2} + c_0 \tilde c)}{\sqrt{3}\tilde c} \frac{x_0 \Lambda}{M_{\rm Pl}},
\end{eqnarray}
\begin{eqnarray}
F_A = c_0 m_{3/2} .
\end{eqnarray}
Here we used $W_0=M_{\rm Pl}/\sqrt{3}$. 
Recall that we will have $F_A \propto x_0^3$ term, which is not suppressed by $M_{\rm Pl}$.
The $R$-axion gets a mass:
\begin{eqnarray}
m_{X_I}^2 = {
2 \mu^4 \over \sqrt 3 x_0 \Lambda M_{\rm Pl}
} = 2\sqrt{3}m_{3/2}^2\left(\frac{M_{\rm Pl}}{x_0 \Lambda}\right)
\end{eqnarray}
at the leading order in $x_0$ and Planck suppressed expansions. 
Now the axino mass is given by
\begin{eqnarray}
 m_{\tilde a} = m_{3/2}+{\eta^2 \tilde c \mu^2 x_0 \Lambda \over f^2}
= m_{3/2}\left(1+\sqrt{3} \eta^2 \tilde c \frac{x_0 \Lambda M_{\rm Pl}}{f^2}\right) .
\end{eqnarray}
The first $m_{3/2}$ contribution vanishes again when we replace the axion kinetic term, $(A+A^\dagger)^2/2$,
with $\exp(A+A^\dagger)$.
\end{itemize}


\section{Solving mixing between $X$ and $A$}
\label{mixing}

\subsection{Kinetic mixing in the standard form by the superfield description: axion mixing and fermion mixing}
Let us consider the following K\"ahler potential, which preserves $U(1)_{\rm PQ}$ symmetry
\begin{eqnarray}
K &=& X^{\dag}X + \frac{f^2}{2} (A+A^{\dag})^2 + K_{XA^{\dag}} X (A+ A^{\dag}) + K_{A X^{\dag}} (A+ A^{\dag}) X^{\dag} .
\end{eqnarray} 
As we want to solve kinetic mixing, we will focus on the vacuum in which $K_{XA^{\dag}}$ has the vev:
\begin{eqnarray}
K \to K_0 = X^{\dag}X + \frac{f^2}{2} (A+A^{\dag})^2 + (A+ A^{\dag})(\kappa X+ \kappa^* X^{\dag}).
\end{eqnarray}
Here we defined $\kappa \equiv \langle K_{XA^{\dag}} \rangle$.
After solving mixing between $A$ and $X$, the above K\"ahler potential becomes
\begin{eqnarray}
K_0 & = & \left( 1- \frac{|\kappa|^2}{f^2}\right) X^{\dag}X + \frac{f^2}{2} (\hat A+ \hat A^{\dag})^2  + ({\rm holomorphic~term}+c.c.).
\end{eqnarray} 
A diagonalized axion $\hat A$ is given by
\begin{eqnarray}
\hat A =  A + \frac{\kappa X}{f^2} = A + \frac{\langle K_{XA^{\dag}} \rangle X}{f^2}.
\end{eqnarray} 
Thus when one considers the canonical normalization we find
\begin{eqnarray}
k\cdot A 
&\to & 
\frac{\sqrt{2}\hat A}{f_a} - \frac{\sqrt{2}X}{f_y} .
\end{eqnarray} 
Here
\begin{eqnarray}
f_a = \sqrt{2}\frac{f}{k}, \ \ \
f_y = \frac{1}{k}\sqrt{2} f \left(\frac{f}{\langle K_{XA^{\dag}} \rangle} \right)\sqrt{1-\frac{|\langle K_{XA^{\dag}} \rangle|^2}{f^2}} .
\end{eqnarray} 
With a gauge kinetic term $S= \frac{1}{2 g^2} + k\frac{A}{8\pi^2}$, we obtain
\begin{eqnarray}
S = \frac{1}{2 g^2} +\frac{1}{8\pi^2}
\left( 
\frac{\sqrt{2}\hat A}{f_a} - \frac{\sqrt{2}X}{f_y}
\right) .
\end{eqnarray}

Furthermore, in the standard form,
$F_X$ and $F$-component of $A$ is given by
\begin{eqnarray}
F_X & \simeq & -(\partial_{X}W)^{\dag} , \\ 
F_A &=& -K^{A X^{\dag}} (\partial_X W)^{\dag}  
\simeq - \frac{K_{X A^{\dag}}}{f^2} F_X .
\end{eqnarray}
Here we used $K_{XX^{\dag}}  \simeq 1, ~ K_{A A^{\dag}} \simeq f^2$ and 
$K_{XX^{\dag}}K_{A A^{\dag}} \gg K_{XA^{\dag}}^2$ for $f \gg K_{XA^{\dag}}$.

Thus, with regard to $f_y$, we can get 
\begin{eqnarray}
f_{y} &\simeq & \frac{1}{k}\sqrt{2} \left(\frac{f^2}{\langle K_{XA^{\dag}} \rangle}\right)  
\simeq  - \frac{1}{k}\sqrt{2} \frac{\langle F_X \rangle}{\langle F_A \rangle } 
 \simeq  \frac{\alpha}{4\pi}\frac{\sqrt{6}m_{3/2}M_{\rm Pl}}{M_{1/2}} .
\end{eqnarray}
Here we used $\langle F_X \rangle \simeq - \sqrt{3}m_{3/2}M_{\rm Pl},$ 
$\langle F_A \rangle \simeq \frac{4\pi}{k\alpha}M_{1/2}$.
Thus we can find
\begin{eqnarray}
{f_a \over f_y} = - {f F^A \over F^X} \equiv -\tan \Theta ,
\end{eqnarray}
where $\Theta$ is a goldstino angle.

Using this goldstino angle, we obtain goldstino $\psi_{X}$ and axino $\tilde a$
\begin{eqnarray}
\psi_{X} \simeq \psi_X^{(0)} - \frac{f_a}{f_y}\psi_A^{(0)} \ \ \
\tilde a \simeq \psi_A^{(0)} + \frac{f_a}{f_y}\psi_X^{(0)} ,
\end{eqnarray}
where we assumed $\Theta \ll 1$.
In the above equation, we normalized fermions canonically and $\psi^{(0)}$ 
means the original field before solving the mixing.
Thus we find $\langle \delta_{\rm SUSY} \tilde a \rangle \sim \langle F_{\hat A} \rangle = 0$.


\subsection{Mixing between $X_R$ and $A_R$}

Remaining fields to be solved are $X_R$ and $A_R$.
Now there is also mass mixing in the scalar potential in addition to the kinetic mixing:
\begin{eqnarray}
V &=& \frac{1}{2}m_{X_R}^2 X_R^2 + \frac{1}{2}m_{A_R}^2 A_R^2 +m_{XA}^2 X_R A_R,
\end{eqnarray}
Here recall that we already solved the kinetic mixing with original field $(X^{(0)}_R, ~A_R^{(0)})$ 
: $X_R = X^{(0)}_R, ~ A_R = A_R^{(0)} + \frac{f_a}{f_y}X^{(0)}_R$.
By an unitary rotation, we will obtain diagonalized fields $(\hat X_R,~\hat A_R)$ with canonical and diagonal kinetic term:
\begin{eqnarray}
V &\approx & \frac{1}{2}m_{X_R}^2(\hat X_R)^2 + \frac{1}{2}m_{A_R}^2 (\hat A_R)^2 .
\end{eqnarray}
Here
\begin{eqnarray}
\nonumber
\begin{pmatrix}
X_R^{(0)} \\
\frac{A_R^{(0)}}{f_a} 
\end{pmatrix}
&=&
\begin{pmatrix}
1 & - \frac{m_{XA}^2}{M^2}f_a  \\
\frac{m_{XA}^2}{M^2 f_a} -\frac{1}{f_y} & 1
\end{pmatrix}
\begin{pmatrix}
\hat X_R  \\
\frac{\hat A_R}{f_a}
\end{pmatrix}
, \ \ 
M^2 = |m_{X_R}^2 -m_{A_R}^2| 
\ \  {\rm for} \ m_{X_R} \neq m_{A_R} , \\
&=& \frac{1}{\sqrt{2}}
\begin{pmatrix}
1 & - f_a  \\
\frac{1}{f_a} -\frac{1}{f_y} & 1 + \frac{f_a}{f_y}
\end{pmatrix}
\begin{pmatrix}
\hat X_R  \\
\frac{\hat A_R}{f_a}
\end{pmatrix}
\ \  {\rm for} \ | m^2_{X_R}-m_{A_R}^2 | < m_{XA}^2 \\
&\equiv &
\begin{pmatrix}
U_{11} & U_{12}  \\
U_{21} & U_{22}
\end{pmatrix}
\begin{pmatrix}
\hat X_R  \\
\frac{\hat A_R}{f_a}
\end{pmatrix}
.
\end{eqnarray}
Here we will always have $m_{X_R} \gtrsim m_{A_R}$ because of $\Lambda_c/\Lambda$; 
hereafter we will not consider the degenerate case.

For Model 0, we will find 
\begin{eqnarray}
\frac{m_{XA}^2}{m_{X_R}^2} 
\sim {\Lambda_c^2  \over fM_{\rm Pl}} 
\sim \left({\Lambda_c \over f}\right)^2 {f_a \over f_y} \leq {f_a \over f_y} .
\end{eqnarray}
Here $f_y \sim M_{\rm Pl}$.


\subsubsection{Mixing between $X_R$ and $A_R$ in Model 1}
In this case, we have
\begin{eqnarray}
m_{X_R}^2  &=& 2\frac{\mu^4}{\Lambda^2} , \ \ \ m^2_{A_R}= 2\eta^2 \tilde c \frac{\mu^4}{f^2}, \\
 m_{XA}^2 & = & \eta^2 \frac{2(-\tilde{d} +\tilde{e} )\mu^4 x_0}{f \Lambda } - \frac{f_a}{f_y}m_{A_R}^2 
 \\
&\sim & km_{X_R}^2 \frac{f_a m_{\tilde a}}{\sqrt{6}m_{3/2}M_{\rm Pl}} 
\left( 1 + \frac{\frac{4\pi}{k\alpha}M_{1/2}}{m_{\tilde a}}\frac{m_{A_R}^2}{m_{X_R}^2}\right)\\
\nonumber
& \simeq & 2.8 \times 10^{12} {\rm GeV}^2
\left(\frac{m_{X_R}}{10^{7}{\rm GeV}}\right)^2
\left(\frac{m_{\tilde a}}{5\times 10^{5}{\rm GeV}}\right) 
\left(\frac{f_{a}}{10^{10}{\rm GeV}}\right)
\left(\frac{30{\rm MeV}}{m_{3/2}}\right) . 
\end{eqnarray}
Here we
supposed $(-\tilde{d} + \tilde{e}) \sim \tilde c$.
Note that mass mixing in the scalar potential $m_{XA}^2$ is of $O(x_0)$
while kinetic mixing $f_y^{-1}$ is of $O(x_0^3)$.

Thus 
\begin{eqnarray}
\frac{m_{XA}^2}{m_{X_R}^2} \simeq k\frac{f_a m_{\tilde a}}{\sqrt{6}m_{3/2}M_{\rm Pl}} 
\simeq 2.8 \times 10^{-2}.
\end{eqnarray}
Here we used used $m_{X_R} \gtrsim m_{A_R}$ and
took $f_a = 10^{10}$ GeV, $m_{\tilde a} = 5\times 10^5$ GeV and $m_{3/2} = 30$ MeV.


\subsubsection{Mixing between $X_R$ and $A_R$ in Model 2}

Recall that eq.(\ref{mxaia}):
\begin{eqnarray}
m_{XA}^2
& = &
  \epsilon_K{\Lambda \over 2 f} {\mu^4 \over f^2} 
\left(
-6 \eta^2\tilde c + {f^2 \over \Lambda^2}
\right) \\
& \equiv & \frac{4\pi}{\alpha}\frac{M_{1/2}f_a}{\sqrt{6}m_{3/2}M_{\rm Pl}}C m_X^2 \\
& \simeq & k\frac{m_{\tilde a}f_a}{\sqrt{6}m_{3/2}M_{\rm Pl}}C m_X^2.
\end{eqnarray}
Here $C=O(1)$, $m_X^2 = \mu^4/\Lambda^2$, and we used $F^A \simeq m_{\tilde a} \simeq \frac{4\pi}{k \alpha}M_{1/2}$.
Then the magnitude of the mass mixing is the same order as that of the kinetic mixing:
\begin{eqnarray}
\frac{m_{XA}^2}{m_X^2}
& \simeq & k\frac{m_{\tilde a}f_a}{\sqrt{6}m_{3/2}M_{\rm Pl}}C  \\ 
&\simeq &  C \frac{f_a}{f_y} 
 \sim  10^{-3} \times C .  
\end{eqnarray}
Here we took $m_{3/2} = 30$ MeV and $m_{\tilde a} \sim 10^4$ GeV.


\subsubsection{Mixing between $X_R$ and $A_R$ in Model $3$}

In this case, we have the similar situation to the Model 1, that is, the effect of the mass mixing is larger than 
that of the kinetic mixing:
\begin{eqnarray}
m_{XA}^2
& \simeq &
- 2 \epsilon \eta^2 \frac{\tilde{d}}{\Lambda f}\mu ^4 \\
& \sim & -2k \frac{m_{\tilde a}}{F^A}\frac{f_a}{f_y} m_X^2  \ \ \ {\rm for} \ \ \ 
\tilde{d} \sim \tilde c
\end{eqnarray}
Then
\begin{eqnarray}
\frac{m_{XA}^2}{m_X^2}
&\sim &  -2 k\frac{m_{\tilde a}f_a}{\sqrt{6}m_{3/2}M_{\rm Pl}} \\
& \simeq & -5.3 \times 10^{-2} .
\end{eqnarray}
Here we took $m_{\tilde a} = 10^5$ GeV and $m_{3/2} = 30$ MeV.


\section{Derivatives of $F^X$ and $F^A$}
\label{derivatives}

We need computations of $\partial_I F^J~(I,J=X,~X^{\dag},~A,~A^{\dag})$ to obtain interactions
between gaugino and those fields for instance.
Now we will focus on the gauge kinetic term:
\begin{eqnarray}
\frac{1}{2}\int d \theta ^2 S W^{\alpha}W_{\alpha} + c.c. , \ \ \
S = \frac{1}{2 g^2} +k\frac{A}{8\pi^2} .
\end{eqnarray}
As we have a gaugino mass from $F^A$,
we can read a coupling of $X$ to gaugino pair from the gaugino mass
at the leading order of fluctuation $\delta_f X$:
\begin{eqnarray}
\frac{1}{2} M_{1/2} &=&  \frac{1}{2}F_A \partial_A \log({\rm Re}(S) (A)) \\
&\simeq & k\frac{\alpha}{4\pi}\frac{K_{X A^{\dag}}}{f^2}(\partial_X W )^{\dag} \\
&=& \frac{1}{2} \langle M_{1/2} \rangle  \left( 1 +  \delta_f X  \langle \partial_X \log\left(K_{XA^{\dag}} \right) \rangle  
\right)+ \cdots \\
& \sim  & \frac{1}{2} \langle M_{1/2} \rangle  
\left( 1 +  \frac{\delta_f X}{\langle X \rangle} \right)  + \dots .
\end{eqnarray}
Here we assumed $K_{XA^{\dag}}$ is a polynomial in $X$, so we can gain
$ \langle \partial_X \log\left(K_{XA^{\dag}} \right)\rangle  \sim  1/\langle X \rangle $ 
unless we have a small $X$ vev or cancellations like Model 2 below.
We used a notation $\langle M_{1/2} \rangle$ to distinguish dynamical fields from a parameter.


\subsection{Derivatives of $F^A$ in Model 0}
\begin{eqnarray}
\frac{1}{3}\partial_X F^X & \simeq & \frac{1}{2} \partial_{X^{\dag}} F^X  \simeq -m_{3/2} , \ \
\partial_{A} F^X =  \partial_{A^{\dag}} F^X 
\sim \frac{f^2}{\Lambda} \frac{m_{3/2}}{M_{\rm Pl}}, \\
\partial_X F^A & \sim &  \frac{m_{3/2}}{M_{\rm Pl}}\left(\frac{m_{A_R}}{m_{X}}\right)^2 \Lambda ,  \ \
\partial_{X^{\dag}} F^A  \sim \frac{m_{3/2}}{M_{\rm Pl}} \Lambda , \ \
\partial_{A} F^A =  \partial_{A^{\dag}} F^A 
\sim m_{3/2}. 
\end{eqnarray}


\subsection{Derivatives of $F^X$ and $F^A$ in Model 1}
\begin{eqnarray}
\partial_X F^X &=&  \partial_{X^{\dag}} F^X 
\sim F^A x_0^2 \left( \eta^2 + \left(\frac{f}{\Lambda} \right)^2 \right), ~~~
\partial_{A} F^X =  \partial_{A^{\dag}} F^X  
\sim F^A x_0 \left(\frac{f}{\Lambda} \right)^2 . \\
\partial_X F^A &=& -\frac{F^A}{x_0}  , ~~~
\partial_{X^{\dag}} F^A  \sim F^A x_0 , ~~~  
\partial_{A} F^A  =  \partial_{A^{\dag}} F^A 
= m_{\tilde a}
\end{eqnarray}
%


\subsection{Derivatives of $F^X$ and $F^A$ in Model 2}
\begin{eqnarray}
\partial_X F^X &=&  \partial_{X^{\dag}} F^X 
\sim F^A \left(\frac{\epsilon_K}{\eta^2}\right)^4 \eta^6 \left(\frac{f}{\Lambda} \right)^4 , ~~~
\partial_{A} F^X =  \partial_{A^{\dag}} F^X 
\sim 
F^A \left(\frac{\epsilon_K}{\eta^2}\right)^5 \eta^6 \left(\frac{f}{\Lambda} \right)^4
\\
\partial_X F^A &\sim & F^A \left(\frac{\epsilon_K}{\eta^2}\right)^3 \eta^8 \left(\frac{f}{\Lambda} \right)^6 , ~~~
\partial_{X^{\dag}} F^A  =  - \epsilon_K ^2 C_3 \frac{ \Lambda ^3 \mu ^2 }{f^4}
\simeq  - \sqrt{2} \Lambda \frac{m_{\tilde a}}{f_y}\frac{1}{k} \\
\partial_{A} F^A  &=&  \partial_{A^{\dag}} F^A
= - C_3 F^A \simeq - m_{\tilde a}.
\end{eqnarray}



\subsection{Derivatives of $F^X$ and $F^A$ in Model 3}
\begin{eqnarray}
\partial_X F^X &=&   {\epsilon  \mu ^2 \over \Lambda}
\sim F^A \left(\frac{f}{\epsilon \Lambda} \right)^2 ,~~~
\partial_{X^{\dag}} F^X = -\frac{\epsilon ^3 \mu ^2}{\Lambda }
\sim F^A \left(\frac{f}{\Lambda} \right)^2
\\
\partial_{A} F^X &=&  \partial_{A^{\dag}} F^X  
= k^2 \frac{f_a^2}{\delta X  \Lambda^2} F^A  \\
\partial_X F^A &=& \frac{1}{2}\partial_{X^{\dag}} F^A  
= \frac{F^A}{\delta X  } ,~~~
\partial_{A} F^A  = \partial_{A^{\dag}} F^A 
= m_{\tilde a}.
\end{eqnarray}



\section{Some decay modes}
\label{decayetc}

Here we will exhibit several decay modes. 
The results can be rough and just show their order of magnitude.


\subsection{The Polonyi and the $R$-axion decay}
\begin{itemize}

\item $\Gamma(X_R \to 2X_I ) $ in Model 1

Through an interaction 
$\frac{X_R}{\sqrt{2}x_0 \Lambda}\partial_{\mu}X_I \partial^{\mu}X_I$ in the K\"ahler potential $K=\Lambda^2 X^{\dag}X$,
we can obtain
\begin{eqnarray}
\Gamma(X_R \to X_I X_I )  
& \simeq & \frac{1}{64 \pi } \frac{m_{X_R}^3}{ (x_0 \Lambda)^2 } .
\end{eqnarray}

\item $\Gamma (X \to 2\psi_{3/2})$

With an interaction $K = \Lambda^2\frac{(X^{\dag}X)^2}{4}$ 
or $K=\Lambda^2 X^{\dag}X$ for the $R$-axion in Model 1\footnote{We have an interaction
between goldstino $\psi_X$ and $R$-axion $X_I$ through $K = \Lambda^{2} X^{\dag}X$:
$\frac{1}{\sqrt{2}x_0 \Lambda}\partial_{\mu} X_I \bar{\psi}_X \bar \sigma^{\mu}\psi_X$. A contribution from 
$K \Lambda^{-2} =\frac{(X^{\dag}X)^2}{4}-\frac{(X^{\dag}X)^3}{18x_0^2}$ will vanish because of the vev $\langle X \rangle \simeq x_0$.}, 
we can find
\begin{eqnarray}
\Gamma (X \to \psi_{3/2}\psi_{3/2}) & \simeq & \frac{1}{96 \pi} \frac{m_{X}^5}{m_{3/2}^2M_{\rm Pl}^2} .
\end{eqnarray}

\item $\Gamma(X \to \psi_{3/2}+ \tilde a )$

Through an interaction $K_{AX}$, we obtain
\begin{eqnarray}
\Gamma (X \to \psi_{3/2}+ \tilde a) & \simeq & \left(\frac{U_{12}}{f_a}\right)^2
\frac{1}{96 \pi} \frac{m_{X}^5}{m_{3/2}^2M_{\rm Pl}^2} .
\end{eqnarray}

\item $\Gamma(X \to 2g)$

Through an interaction with mixing between $A$ and $X$,
\begin{eqnarray}
\frac{\alpha}{8\pi} \left( 
\frac{U_{12}}{f_a} \frac{X_R}{f_a} F_{\mu \nu}F^{\mu \nu} 
+ \frac{f_a}{f_y} \frac{X_I}{f_a} F_{\mu \nu}\tilde{F}^{\mu \nu}
\right),
\end{eqnarray}
we obtain
\begin{eqnarray}
\Gamma(X_R \to g g)
&\simeq &
\frac{ N_g}{16 \pi} \left(\frac{\alpha}{4\pi}\right)^2 \left(\frac{U_{12}}{f_a}\right)^2 
\frac{m_{X_R}^3}{f_a^2} \\ 
\Gamma(X_I \to g g)
&\simeq &
\frac{ N_g}{16 \pi} \left(\frac{\alpha}{4\pi}\right)^2 \left(\frac{f_a}{f_y}\right)^2 
\frac{m_{X_I}^3}{f_a^2} .
\end{eqnarray}
For Model 0, we will not have a loop factor.

\item $\Gamma(X \to 2\tilde a )$ 

Through an interaction
\begin{eqnarray}
& & m_{\tilde a} \frac{X}{\sqrt{2}\langle X \rangle}\tilde a \tilde a \ \ {\mbox{for Model }} 1,3 , \\ 
& & m_{\tilde a}  \frac{X}{kf_y}\tilde a \tilde a \left( 1+ \frac{C_3}{2}\left(\frac{m_{A_R}}{m_{\tilde a}} \right)^2\right)
{\mbox{for Model }} 2,0 
\end{eqnarray}
we obtain
\begin{eqnarray}
\Gamma(X \to \tilde a \tilde a)
&\simeq &
\frac{1}{32 \pi} m_{X} \left(\frac{m_{\tilde a}}{\langle X \rangle} \right)^2 \ \ {\mbox{for Model }} 1,3 , \\ 
&\simeq &
\frac{1}{16 \pi} m_{X} \left(\frac{m_{\tilde a}}{kf_y} \right)^2
\left( 1+ \frac{C_3}{2}\left(\frac{m_{A_R}}{m_{\tilde a}} \right)^2\right)^2 \ \ \ {\mbox{for Model }} 2,0 .
\end{eqnarray}
Note that $\langle X \rangle$ is a dimensionful vev of $X$ and
the above interactions are obtained via $\eta^2 \Lambda^2 \frac{\tilde c}{2} (A+A^{\dag})^2 X^{\dag} X$ and 
$\frac{f^2 C_3}{3!} (A+A^{\dag})^3$.

\item $\Gamma(X \to 2\lambda )$

For a mode $X \to \lambda \lambda$, $\partial_X F^A$ or $\partial_{X^{\dag}} F^A$ becomes a coupling of $X$ to gaugino pair:
\begin{eqnarray}
\Gamma(X \to \lambda \lambda)
&\simeq &
\frac{N_g}{32 \pi} m_{X} \left(\frac{M_{1/2}}{\langle X \rangle} \right)^2 \ \ {\mbox{for Model }} 1,3 , \\ 
&\simeq &
\frac{C_3^2}{16 \pi k^2} m_{X} \left(\frac{M_{1/2}}{f_y} \right)^2 \ \ \ {\mbox{for Model }} 2,0 .
\end{eqnarray}
Thus a partial decay width of this mode can be suppressed by 
$\left(\frac{M_{1/2}}{m_{\tilde a}}\right)^2$ for Model 1 and 3
or $\left(\frac{M_{1/2}}{m_{\tilde a}}\right)^2 \left( 1+ \frac{C_3}{2}\left(\frac{m_{A_R}}{m_{\tilde a}} \right)^2\right)^{-2}$ 
for Model 2 and 0, comparing with $\Gamma(X \to \tilde a \tilde a)$.
If we have $K= c_{\mu}(A+A^{\dag})H_u H_d + c_B (A+A^{\dag})^2 H_u H_d$,
$\Gamma(X \to \tilde h_u \tilde h_d)$ would be comparable to this mode, supposing 
$c_{\mu} \sim 10^{-2}$, $c_{B} \lesssim 10^{-4}$ and $\mu \sim M_{1/2}$.

\item $\Gamma_{X_I}$ in Model 1

For Model 1, we find through an interaction $-i \frac{X_I}{\sqrt{2}x_0 \Lambda} \frac{M_{1/2}}{2 }\lambda \lambda  +c.c. $
\begin{eqnarray}
\Gamma (X_I \to \lambda \lambda) &=&
\frac{N_g}{32\pi} m_{X_I} \left(\frac{M_{1/2}}{\Lambda x_0}\right)^2 \sqrt{1- \frac{4M_{1/2}^2}{m_{X_I}^2}} .
\end{eqnarray}
For $m_{X_I} < 2M_{1/2}$,
the channel $X_I \to \lambda \lambda$ is closed, then $X_I \to b \bar{b}$ can become the main decay mode: 
\begin{eqnarray}
\Gamma (X_I \to b \bar b) &\simeq &
\frac{3m_{X_I}}{16 \pi}\left(
\frac{m_A^2 \sin^2 \beta}{x_0 \Lambda}\frac{m_{b}}{m_A^2-m^2_{X_I}}
\right)^2 \sqrt{1-{4m_{b}^2 \over m_{X_I}^2}} .
\end{eqnarray} 
For a mode of $X_I \to \tau \bar{\tau}$, $m_b$ should be replaced with $m_{\tau}$.

\item $\Gamma( X_R \to 2\Phi)$

We denoted MSSM scalar fields as $\Phi$.
For decay mode $X_R \to 2\Phi$, 
$K= k\left(\frac{\alpha}{4\pi}\right)^2 (A+A^{\dag})^2 \Phi^{\dag}\Phi$ is important.
These decay occur via soft scalar mass 
$\frac{m_0^2}{\langle X \rangle} X_R \Phi^{\dag}\Phi$ or $\frac{m_0^2}{f_y} X_R \Phi^{\dag}\Phi$.
These amplitude is given by 
\begin{eqnarray}
\Gamma(X_R \to 2\Phi) \sim \frac{N_m}{N_g}\left(\frac{m_0}{M_{1/2}}\right)^2 \left(\frac{m_0}{m_{X_R}}\right)^2
\times \Gamma (X \to \lambda \lambda). 
\end{eqnarray} 
Here $N_m$ is the number of these decay channels.
Supposing we have
$K= c_{\mu}(A+A^{\dag})H_u H_d + c_B (A+A^{\dag})^2 H_u H_d$ and
$B \mu \sim m_0^2$, 
$\Gamma(X \to H_u H_d)$ from this interaction would be comparable to this mode.
The decay may also occur via derivative interaction
$k\left(\frac{\alpha}{4\pi}\right)^2 \langle A_0 \rangle \left( \frac{U_{12}}{f_a}\right) 
\frac{X_R}{f_a}\Phi^{\dag}\partial^2 \Phi$ \cite{Moroi:1999zb}.
(Note that $A$ includes $\langle A_0 \rangle$.)
However the effect of this interaction is much smaller than the above result. 
For Model 0, we will have similar result to the Model 2.

\item $\Gamma( X_R \to  2 a )$ etc.

Through an interaction
\begin{eqnarray}
\frac{U_{12}}{f_a}\frac{X_R}{kf_a} \partial_{\mu}a\partial^{\mu}a
&\simeq & \frac{m_{\tilde a}}{\sqrt{6}m_{3/2}M_{\rm Pl}} X_R \partial_{\mu}a\partial^{\mu}a ,
\end{eqnarray}
we will obtain
\begin{eqnarray}
\Gamma(X_R \to a  a)
&\simeq &
\frac{1}{192 \pi}  \frac{m_{X_R}^3 m_{\tilde a}^2}{m_{3/2}^2M_{\rm Pl}^2} . 
\end{eqnarray}
Note that the above interactions are obtained via 
$\eta^2 \Lambda^2\frac{\tilde c}{2} (A+A^{\dag})^2 X^{\dag} X$ and $\frac{f^2 C_3}{3!} (A+A^{\dag})^3$
for Model 1,~3 
or via $\frac{f^2 C_3}{3!} (A+A^{\dag})^3$ for Model 2.
For $m_X \gg m_{A_R}$, we have the same partial decay width of $X_R \to A_R A_R$ as this.


\subsection{Saxion decay}

\item $\Gamma (A_R \to 2a )$

Through an interaction 
$\hat{C}_3\frac{A_R}{f_a}\partial_{\mu} a \partial^{\mu} a$ in the K\"ahler potential 
$\frac{C_3 f^2}{3!} (A+A^{\dag})^3$, we can compute
\begin{eqnarray}
\Gamma (A_R \to a + a ) 
&= & \frac{1}{32 \pi k^2} {C}_3^2 \frac{m_{A_R}^3}{f_a^2} .
\end{eqnarray}

\item $\Gamma (A_R \to a + X_I )$

From the previous interaction, we have
\begin{eqnarray}
\Gamma (A_R \to a + X_I ) 
&= & \frac{1}{32 \pi k^2}{C}_3^2 \left(\frac{f_a}{f_y}\right)^2\frac{m_{A_R}^3}{f_a^2} .
\end{eqnarray}

\item $\Gamma(A_R \to \psi_{3/2}+ \tilde a)$

Through an interaction $K=\eta^2\frac{\tilde c}{2}(A+A^{\dag})^2 X^{\dag}X$, we can find
\begin{eqnarray}
\Gamma(A_R \to \psi_{3/2}\tilde a) &\simeq &
\frac{1}{96 \pi } \frac{m_{A_R}^5}{m_{3/2}^2 M_{\rm Pl}^2} .
\end{eqnarray}

\item $\Gamma(A_R \to 2 \psi_{3/2})$

Through a mixing between $X_R$ and $A_R$, we can find
\begin{eqnarray}
\label{saxiongravitino}
\Gamma(A_R \to \psi_{3/2}\psi_{3/2}) &\simeq &
\frac{1}{48 \pi } \left( \frac{U_{12}}{f_a}\right)^2 \frac{m_{A_R}^5}{m_{3/2}^2 M_{\rm Pl}^2} .
\end{eqnarray}
But this will be suppressed by $\left( \frac{U_{12}}{f_a}\right)^2$, compared to $\Gamma(A_R \to \psi_{3/2}+ \tilde a)$.

\item $\Gamma(A_R \to 2g )$

Through an interaction
\begin{eqnarray}
\frac{\alpha}{8\pi} \frac{A_R}{f_a} F_{\mu \nu}F^{\mu \nu} 
\end{eqnarray}
we obtain
\begin{eqnarray}
\Gamma(A_R \to g g)
&\simeq &
\frac{N_g}{16 \pi} \left(\frac{\alpha}{4\pi}\right)^2 \frac{m_{A_R}^3}{f_a^2} .
\end{eqnarray}
For Model 0, we do not have a loop factor in terms of $f$.

\item $\Gamma(A_R \to 2\tilde a)$ 

Through an interaction
\begin{eqnarray}
{C}_3 m_{\tilde a}\frac{A_R}{k f_a}\tilde a \tilde a
\end{eqnarray}
in $\frac{f^2C_3}{3!}(A+A^{\dag})^3$,
we will obtain
\begin{eqnarray}
\Gamma(A_R \to \tilde a \tilde a)
&\simeq &
\frac{C_3^2}{4 \pi k^2}m_{A_R} \left(\frac{m_{\tilde a}}{f_a} \right)^2 \sqrt{1-4\frac{m_{\tilde a}^2}{m_{A_R}^2}}.
\end{eqnarray}
Here we used $\partial_{A_R} F^A \simeq 2m_{\tilde a}$.
Note that we may have also an interaction $\eta^2 \frac{d'}{2k} m_{\tilde a}\frac{A_R}{f_a}\tilde a \tilde a$ via
$K\Lambda^{-2} = -\eta^2 \frac{\tilde c}{2} (A+A^{\dag})^2 X^{\dag}X $, 
where $ d' $ is a coefficient of mixing matrix $U_{12}$, e.g.,
$d' = \eta^2(\tilde{d} - \tilde{e})$ for Model 1.
But this is at most comparable to the above interaction.

\item $\Gamma(A_R \to 2\lambda)$ 

Note a decay mode $A_R \to \lambda \lambda$ comes from a gaugino mass interaction 
$\frac{\alpha}{4\pi}m_{\tilde a} \frac{A_R}{f_a}\lambda \lambda$ with using $\partial_{A_R} F^A \simeq 2 m_{\tilde a}$.
Thus 
\begin{eqnarray}
\Gamma(A_R \to \lambda \lambda)
&\simeq &
 N_g \left(\frac{\alpha}{4\pi}\right)^2
\frac{m_{A_R}}{4 \pi }\left(\frac{m_{\tilde a}}{f_a} \right)^2 \sqrt{1-4\frac{m_{\tilde a}^2}{m_{A_R}^2}}.
\end{eqnarray}
is suppressed by $N_g \left(\frac{\alpha}{4\pi}\right)^2$,
compared to $\Gamma(A_R \to \tilde a \tilde a)$.
Note that $\Gamma(A_R \to \tilde h_u \tilde h_d)$ would be comparable to this mode, supposing 
we have $K= c_{\mu}(A+A^{\dag})H_u H_d + c_B (A+A^{\dag})^2 H_u H_d$,
$c_{ \mu} \sim \frac{\alpha}{4\pi}$.
For Model 0, $k\frac{\alpha}{4\pi}m_{\tilde a}$ should be replaced by $m_{3/2} \sim M_{1/2}$.

\item $\Gamma(A_R \to 2\Phi)$

For a decay mode $A_R \to 2\Phi$, 
$K= k\left(\frac{\alpha}{4\pi}\right)^2 (A+A^{\dag})^2 \Phi^{\dag}\Phi$ is important.
This decay can occur via soft scalar mass
$\frac{\alpha}{4\pi}\frac{m_{\tilde a}m_0}{\sqrt{k}f_a} A_R \Phi^{\dag}\Phi$. 
The amplitude is given by 
\begin{eqnarray}
\Gamma (A_R \to 2\Phi) \sim \frac{N_m}{N_g}
\left(\frac{m_0}{m_{A_R}}\right)^2
\times \Gamma (A_R \to \lambda \lambda) .
\end{eqnarray}
The decay may also occur via derivative interaction
$\left(\frac{\alpha}{4\pi}\right)^2 \langle A_0 \rangle \frac{A_R}{f_a}\Phi^{\dag}\partial^2 \Phi$.
However, this effect is much smaller than the above result.
For Model 0, we will have a similar result, except that an interaction $\frac{A_R}{f}m_{0}^2 \Phi^{\dag}\Phi$ becomes relevant.

\item $\Gamma(A_R \to 2X_I )$ in Model 1  

Note that we have $\frac{U_{12}}{\tilde{d}  \langle X \rangle} \sim \eta^2$ for Model 1.
(Recall that $\langle X \rangle$ is a dimensionful parameter.)
Then we have interaction
$\frac{U_{12}}{\langle X \rangle} 
\frac{A_R}{f_a}\left(\partial_{\mu}X_I \partial^{\mu}X_I + \delta \partial_{\mu}X_R \partial^{\mu}X_R\right)$ for Model 1.
Thus, through this interaction we can find
\begin{eqnarray}
\Gamma(A_R \to X_I X_I)
&\sim &
\frac{1}{64 \pi} \left(\frac{U_{12}}{\langle X \rangle}\right)^2 \frac{m_{A_R}^3}{f_a^2}  \ \ {\rm for ~Model} ~1~X_I .
\end{eqnarray}
Here $\frac{U_{12}}{\langle X \rangle} \sim \eta^2$ for Model 1.


\subsection{Axino decay}
\label{axinodecay}

\item $\Gamma(\tilde a \to \lambda + g)$

Through an interaction in gauge kinetic term
\begin{eqnarray}
\frac{\alpha}{16\pi f_a} \tilde a \sigma^{\mu \nu}\lambda^a F^a_{\mu \nu} + c.c. ,
\end{eqnarray}
the heavy axion mainly decays into gluino and gluon because of the strong interaction. Then we have 
\begin{eqnarray}
\label{axino1}
\Gamma(\tilde a \to \lambda + g) 
&\simeq &
N_g \frac{\alpha^2}{256 \pi^3} \frac{m_{\tilde a}^3}{f_a^2} 
\end{eqnarray}
For Model 0, there is no loop factor in terms of $f$.

\item $\Gamma(\tilde a \to \psi + \Phi)$

Here we denoted $\psi$ as the MSSM matter fermions.
Note that we also have the fermion-sfermion-axino interaction
\begin{eqnarray}
\frac{\alpha^2}{\sqrt{2}\pi^2}\frac{M_{1/2}}{f_a}\log\left(\frac{f_a}{M_{1/2}}\right)
\Phi \psi \tilde a
\end{eqnarray}
via loop correction by the MSSM gauge interactions \cite{Covi:2004rb}. 
However, this interaction is irrelevant since axino mass is large: amplitude 
$\Gamma(\tilde a \to  \psi + \Phi )$
will be suppressed by $ {N_m \over N_g} \left(\alpha \log\left(\frac{f_a}{M_{1/2}}\right)\right)^2(M_{1/2}/m_{\tilde a})^2$
compared to $\Gamma(\tilde a \to \lambda + g)$.

\item $\Gamma(\tilde a \to \psi_{3/2} + a)$

Through an interaction
\begin{eqnarray}
\frac{i}{M_{\rm Pl}}\tilde a  \sigma^{\mu} \bar\sigma^{\nu} \psi_{\mu} \partial_{\nu} a
\end{eqnarray}
or equivalently,
\begin{eqnarray}
i \frac{m_{\tilde a}}{\sqrt{6}m_{3/2}M_{\rm Pl}} \tilde a \sigma^{\mu} \bar \psi_X \partial_{\mu} a ,
\end{eqnarray}
this decay will occur. Here the above goldstino interaction originate from 
$-\eta^2 \Lambda^2 \frac{\tilde c}{2} (A+A^{\dag})^2 |X|^2$ or $\frac{C_3 f^2}{3!}(A+A^{\dag})^3$. 
Then we find
\begin{eqnarray}
\label{axino2}
\Gamma(\tilde a \to \psi_{3/2} + a) 
&\simeq &
\frac{1}{96 \pi} \frac{m_{\tilde a}^5}{m_{3/2}^2 M_{\rm Pl}^2} 
\end{eqnarray}

\item $\Gamma(\tilde a \to \psi_{3/2} + X_I)$ in Model 1

In Model 1, we can have an $R$-breaking interaction
$m_{\tilde a}\frac{X_I}{x_0 \Lambda}\frac{f_a}{f_y}\tilde a \psi_X$ in $K_{AX}$. Thus we find
\begin{eqnarray}
\Gamma(\tilde a \to \psi_{3/2} + X_I) 
&\simeq &
\frac{1}{8 \pi} \left(\frac{f_a}{f_y}\right)^2\frac{m_{\tilde a}^3}{(x_0\Lambda)^2} .
\end{eqnarray}

\end{itemize}

\end{document}